\documentclass[aip,jcp,preprint,superscriptaddress]{revtex4-1}

\usepackage{graphicx}
\usepackage{amsmath}
\usepackage{bm}
\usepackage{txfonts}

\def\x1{{\bs x}_1}              %
\def\x2{{\bs x}_2}              %
\def\x3{{\bs x}_3}              %
\def\wi{\mathrm{Wi}}              %
\def\gam{\dot{\gamma}}

\def\omegal{\omega}              %
\def\omegae{\Omega}              %
\def\omegat{\Omega_B}              %
\def\omegar{\omega_{R}}              %

\begin{document}


\title{Application of the Eckart frame to soft matter: rotation of star polymers under shear flow}


\author{Jurij Sabli\'{c}}
\affiliation{Department of Molecular Modeling, National Institute of Chemistry, Hajdrihova 19, SI-1001 Ljubljana, Slovenia}

\author{Rafael Delgado-Buscalioni}
\email[]{rafael.delgado@uam.es}
\affiliation{Departamento F\'{i}sica Te\'{o}rica de la Materia Condensada, Universidad Aut\'{o}noma de Madrid, Campus de Cantoblanco, E-28049 Madrid, Spain}
\affiliation{Condensed Matter Physics Center, IFIMAC, Campus de Cantoblanco, E-28049 Madrid, Spain}

\author{Matej Praprotnik}
\email[]{praprot@cmm.ki.si}
\affiliation{Department of Molecular Modeling, National Institute of Chemistry, Hajdrihova 19, SI-1001 Ljubljana, Slovenia}
\affiliation{Department of Physics, Faculty of Mathematics and Physics, University of Ljubljana, Jadranska 19, SI-1000 Ljubljana, Slovenia}





\begin{abstract}
The Eckart  co-rotating frame is used to  analyze the  dynamics of star polymers under shear flow, either in melt or solution and with different types of bonds.  This formalism  is compared  with the  standard approach used in many previous  studies on  polymer  dynamics, where  an apparent  angular velocity $\omegal$ is obtained from relation between the tensor of  inertia and angular  momentum.  A common mistake is to interpret $\omegal$  as the molecular rotation  frequency, which is only valid  for rigid-body  rotation.  The Eckart  frame, originally formulated to analyze the infrared spectra of small molecules, dissects different  kinds of  displacements:  vibrations without  angular momentum,  pure  rotation,  and vibrational  angular momentum (leading to a Coriolis  cross-term). The  Eckart frame co-rotates   with   the   molecule with   an   angular   frequency $\omegae$  obtained  from  the Eckart  condition  for  minimal coupling between  rotation and  vibration.  The standard  and Eckart approaches are compared with a  straight description of the star's dynamics  taken from  the  time autocorrelation  of the monomers positions moving around  the molecule's center of mass.   This is an underdamped oscillatory signal, which can  be described by a rotation frequency  $\omegar$   and  a  decorrelation  rate   $\Gamma$.   We consistently  find that  $\omegae$  coincides with  $\omegar$, which determines the characteristic tank-treading rotation of the star. By  contrast,  the apparent  angular velocity  $\omegal<\omegae$  does not  discern  between  pure rotation  and molecular  vibrations. We believe  that the Eckart frame will be useful to unveil the dynamics of semiflexible molecules  where  rotation and deformations  are  entangled, including tumbling, tank-treading motions and breathing modes.
\end{abstract}


\maketitle


\section{Introduction}

Soft  matter  and  in  particular,  polymers,  exhibit  quite  rich
dynamics under  non-equilibrium conditions.  A plethora  of collective
motions has been  described in the literature, not  only of polymers,
but also  of vesicles  and more.   In a shear  flow, the  steady state
conformation  is  not  possible,  and  linear  polymers  perform  wild
conformational  changes  stretching   and  tumbling  \cite{Smith:1999, Teixeira:2005,tumbling, Costanzo:2016, Winkler:2016, Chen:2017}. Star   molecules,  dendrimers   and  also
vesicles face the shear flow in a different way. They perform internal
rotations around the  molecule center of mass (CoM),  while keeping their
overall shape  and orientation  roughly fixed.   This motion  has been called
tank-treading \cite{Abkarian:2007,Alireza:2011, Dodson:2010, Aust:2002, Ripoll:2006, Ripoll:2007, Chen:2017}. The  case of  ring polymers, whose properties have recently been extensively studied~\cite{Yan:2016,  Hsiao:2016, Yoon:2016, Chen:2017},  is probably  in between  and recent  work
indicates that they tumble or tank-tread depending on the value of the
shear rate~\cite{Chen:2013, Chen:2015:1,
  Chen:2015:2}.  Recently, we have observed that star molecules
under large  enough shear flow perform  another collective oscillation,
with successive  extensions and  contractions in their  overall length.
We called this mode ``breathing'' \cite{Sablic:2016:1} and showed that its
characteristic frequency $\omegat$ has the same physical origin as the
tumbling frequency  in linear  and ring  chains. The  difference being
that soft  stars do not tumble,  but rather let their  arms rotate.  A
similar breathing mode (probably  with different mechanical origin) is
also  observed in  vesicles \cite{Abkarian:2007,Alireza:2011, Dodson:2010}.
It could, however,  well be that star  molecules with
attractive inter-monomer  interactions (i.e. in bad  solvent) would not only
 tank-tread,  but also  occasionally tumble  (a rotation  of the
overall molecular shape)  like a rugby ball does.  One could speculate
that  the  stiffer  the  intermonomer interactions,  the  larger  the
resemblance with a  rigid body would be; with  some dynamic transition
(tank-tread-to-tumble) taking place  at moderate attractive energies.
Would still those semi-rigid stars breath?  These sort of questions on
the mechanics of soft deformable macromolecules are difficult to study
in  clean  ways.   The reason  is  clear:  at  a  given time,  in  the
laboratory (inertial)  frame, it  is not  possible to  discern between
pure rotations and  vibrations of the molecule. The  simple shear flow
is a paradigm of  such duality because it is a mix  of a pure rotation
and a pure  shear strain (stretching in one  direction and compressing
over the  perpendicular line).   Hybrid affine  deformations combining
shear   and   pure elongational   flows   have also been   studied
\cite{Jain:2015},  enriching    the   dynamic    panorama.    The
complications of using  the laboratory frame to study  the rotation of
non-rigid molecules have been overlooked  in many previous works on polymer
dynamics. In  particular, a simple estimation of  the molecular angular  velocity $\omegal$ based  on the polymer shape was first proposed in Ref.~\cite{Cerf:1969}. Such relation, stems  from the rotation dynamics in the
laboratory frame, where the angular velocity $\omegal$ is related with
the total angular  momentum ${\bf L}$ and inertia tensor  ${\bf J}$ as
${\bf L}= {\bf J}\cdot {\bm \omegal}$.  A particularly simple estimations of
$\omegal$ involving  the gyration  tensor components, was  proposed as
rotational-optic rule \cite{Cerf:1969, Aust:2002}.  Since,
many works have  reported values of $\omegal$ and used it to interpret
the polymer  rotational dynamics, this
has led to  erroneous interpretations which is still  very much alive
in the literature \cite{Ripoll:2006,Singh:2012,Singh:2013,Yamamoto:2015,Xu:2016, Aust:2002}.

We have recently completed a series of works on star polymer
dynamics\cite{DelgadoBuscalioni:2015,Sablic:2016,Sablic:2016:1}.
This  series started  by  a study  of the  effect  of open  boundaries
compared  with closed  systems in  the rheology  of melts  under shear
(simulations using  Open Boundary  Molecular Dynamics (OBMD)
\cite{hmd_prl06,DelgadoBuscalioni:2015,Sablic:2016} permits to fix the  pressure load and shear stress,
instead of the density and shear  velocity). As a continuation of such
work, we  studied   the  dynamics  of  stars  in   solution  and  melt
\cite{Sablic:2016:1}  and observed  that the  tank-treading frequency  of
monomers around  the molecule's  CoM, $\omegar$, was  completely different
from  the ``apparent''  angular  velocity obtained  from the  standard
(lab-frame)  analysis  $\omegal$.  We  also  noted  that  the origin of such  strong differences was not explained in previous works.  Motivated by
these observations, we decided to  tackle the problem of soft molecule
rotational  dynamics   using  an  old  and   robust  formalism,  which
apparently, has been  largely forgotten by the  soft matter community:
the Eckart co-rotating frame.

The  Eckart frame formalism, derived in 1935  \cite{Eckart:1935}, uses a non-inertial
frame, which rotates with the molecule.
It allows to disentangle translation, rotations, and  vibrations.
Aside from vibrations without angular momentum contribution (which
can be detected in the inertial frame), the non-inertial frame
allows to reveal vibrations with angular momentum. These are the
displacements with respect to a purely rotating (rigid-body) reference configuration.
The Eckart condition determines the rotation frequency of the reference configuration
by minimizing the coupling between vibrational angular momentum and pure rotation~\cite{Wilson:1955}.
The calculus of the so called ``Eckart angular velocity'' $\omegae$,
can    be    carried    out   by    the    Eckart    frame formalism and has been mostly used to study
the Raman spectra of small molecules
\cite{Louck:1976, Rhee:1997}
as well as in a variety of other applications, such as structural isomerization dynamics of atomic clusters \cite{Yanao:2004} or molecular dynamics (MD) integration\cite{Praprotnik:2005, Praprotnik:2005:1, Praprotnik:2005:2, Praprotnik:2005:3}. The ``apparent''  angular  velocity $\omegal$
extracted from the total angular momentum in the inertial frame, mixes up pure rotation
and vibrational angular momentum. A misinterpretation of this apparent angular velocity
had as consequence some large discrepancies in the polymer literature on shear flow \cite{Ripoll:2006, Chen:2013}.

While  the  Eckart  formalism  is traditionally  used  in  equilibrium
states, here we use it to  describe a situation which is far-away from
equilibrium.  Although  the Eckart condition is  first-order accurate,
we show  that it is robust  enough to capture the  correct physics. In
particular,  we show  that  the  Eckart frame  is  independent on  the
reference configuration chosen (see Appendix)  and that, for any shear
rate, the resulting  frequency $\omegae$ equals within  error bars the
monomer rotation  frequency about  the molecule CoM,  $\omegar$.  Star
polymers are particularly  interesting for this sort  of study because
of their rich dynamics in shear flow (with tank-treading and breathing
modes \cite{Sablic:2016:1})  and also because they  represent a bridge
between  the   physics  of  polymers   and  colloids~\cite{Likos:1998,
  Likos:2001, Grest:2007}.  More generally,  we expect this  work will
foster the use of Eckart frame  as another useful tool in the analyses
of flowing macromolecules' dynamics.

The  remainder  of the  paper  is  structured  as follows:  first,  we
describe the standard (laboratory frame) analysis and the Eckart frame.
Then, we describe our working models (star polymer in melt and solution under shear flow). Results and discussion are then presented, followed by conclusions.

\section{Dynamics description in the laboratory frame}
A standard approach to describe the rotation of molecules
is based on the inertial frame (laboratory frame)
and follows from a straight generalization of the rigid body rotation,
allowing for vibrations without angular momentum contribution $\tilde{\bf v}$.
The kinetic equation for the time evolution of the position of the $\alpha$ monomer
${\bf r}_{\alpha}$ is,
\begin{equation}
\dot{\mathbf{r}}_{\alpha} = \dot{\mathbf{r}}_{cm} + {\bm \omegal} \times \left( \mathbf{r}_{\alpha} - \mathbf{r}_{cm} \right) + \mathbf{\tilde{v}}_{\alpha}.
\label{eqn:velocity monomer standard}
\end{equation}  
In the standard (lab frame) description, the vibrational motion
is angular momentum free, and it is denoted  by $\mathbf{\tilde{v}}_{\alpha}$.
It is particularly strong in soft molecules as polymers.
The corresponding angular frequency is then ~\cite{Ripoll:2006, Aust:2002, Xu:2016}
\begin{equation}
{\bm \omegal} = \mathbf{J}^{-1} \cdot\mathbf{L}.
\label{eqn:standard eqn. omega}
\end{equation}  
Here, $\mathbf{L} = \sum_{\alpha=1}^{N} \left(\mathbf{r}_{\alpha} - \mathbf{r}_{cm}\right) \times m_{\alpha}\left(\mathbf{v}_{\alpha} - \mathbf{v}_{cm}\right)$
is the angular momentum of the rotating molecule and $\mathbf{J}$
its moment-of-inertia tensor with respect to the position of its CoM $\mathbf{r}_{cm}$,
defined as
\begin{equation}
\begin{split}
 \mathbf{J} = \sum_{\alpha=1}^{N} m_{\alpha} \lbrace \left[\left(\mathbf{r}_{\alpha} - \mathbf{r}_{cm}\right) \cdot \left(\mathbf{r}_{\alpha} - \mathbf{r}_{cm}\right) \right]\bm{\mathbf{I}} - \\
\left(\mathbf{r}_{\alpha} - \mathbf{r}_{cm}\right) \otimes \left(\mathbf{r}_{\alpha} - \mathbf{r}_{cm}\right) \rbrace ,
\label{eqn:standard inertia tensor}
\end{split}
\end{equation}
with $\mathbf{I}$ being a $3\times 3$ identity matrix, $\mathbf{r}_{\alpha}$
the coordinate vector of monomer $\alpha$ of the molecule, and $m_{\alpha}$ its mass
(here $m_{\alpha}=1$).

A common mistake is to interpret $\omegal$ as the molecular angular velocity.
However, $\omegal$ does not describes the pure rotational component of the molecule
and in fact, it is called the {\em apparent angular velocity}
in the literature dealing with the Eckart formalism \cite{Rhee:1997}.
Only in the case of rigid-body
motion ($\tilde{v}=0$) does $\omegal$ coincides with the rotational angular velocity.
The reason will come clear in the next section.

\section{Description using the co-rotating Eckart frame}
The Eckart formalism permits to dissect yet
another kind of vibrations ${\bf u}$, which
contribute to the total angular momentum,
but {\em do not contribute to the molecular rotation frequency}.
The  Eckart frame  is a non-inertial frame, which
co-rotates with the molecule attached to its CoM.
The {\em pure rotation} frequency $\omegae$ is obtained by minimizing
the coupling between pure rotation and this vibrational angular momentum:
the Coriolis  coupling is minimal in this internal moving frame
\cite{Eckart:1935,  Wilson:1955,  Louck:1976}.  

The  first step  of  the Eckart  frame  formalism  is  to choose some
{\em rigid} molecular configuration, which is taken as the reference one~\cite{Louck:1976}.
The Eckart frequency and kinetic energy are, however, independent
on the {\em rigid} reference configuration chosen.
This fact is illustrated in the Appendix, where we compare three  different reference  configurations.
Once the reference configuration is chosen, we introduce  the initial internal coordinate
system,   defined    by   the   three   right-handed    base   vectors
$\mathbf{f}_{1}$,  $\mathbf{f}_{2}$,  and  $\mathbf{f}_{3}$  with  the
origin in  the CoM  of the molecule.  The initial  internal coordinate
frame  $\left(\mathbf{f}_{1},  \mathbf{f}_{2},  \mathbf{f}_{3}\right)$
can   be  chosen   arbitrarily,  i.e.   its  initial   orientation  is
arbitrary. The  components of the  position vector of  the $\alpha$-th
monomer  of  the  reference  configuration expressed  in  the  initial
internal    coordinate   system    are   denoted    as   $c_i^\alpha$,
$i=1,2,3$. Once defined, the  $c_i^\alpha$s remain constant during the
computation    of   the    angular    velocity    and   fulfill    the
equation~\cite{Eckart:1935, Louck:1976}:
\begin{equation}
\sum _{\alpha=1}^{N} m_{\alpha} c_i^\alpha = 0, \,\,\,\,\mathrm{for}\;i=\left\{1,2,3\right\}.
\end{equation}  
The CoM velocity in thus defined internal coordinate frame is $0$~\cite{Eckart:1935, Louck:1976}. From the instantaneous positions of the monomers in the polymer and with the $c_i^\alpha$s, we define the three Eckart vectors $\mathbf{\mathcal{F}}_{1}$, $\mathbf{\mathcal{F}}_{2}$, and $\mathbf{\mathcal{F}}_{3}$, which are given by~\cite{Eckart:1935, Louck:1976}:
\begin{equation}
\mathbf{\mathcal{F}}_{i} = \sum_{\alpha=1}^{N} m_{\alpha}c_i^{\alpha}\left( \mathbf{r}_{\alpha} - \mathbf{r}_{cm} \right).
\label{eqn:Eckart_vectors}
\end{equation} 
From the Eckart vectors, we define a symmetric positive definite Gram matrix $\mathcal{F}$ with the ${ij}$-component defined as $\left[\mathcal{F}\right]_{ij} = \mathcal{F}_{i} \cdot \mathcal{F}_{j}$. The unit base vectors of the instantaneous Eckart frame,
(defined by the instantaneous positions of monomers) are computed as~\cite{Eckart:1935, Louck:1976}:
\begin{equation}
\left( \mathbf{f}_{1}, \mathbf{f}_{2}, \mathbf{f}_{3} \right) = \left(\mathbf{\mathcal{F}}_{1}, \mathbf{\mathcal{F}}_{2}, \mathbf{\mathcal{F}}_{3}\right) \, \mathcal{F}^{-1/2}.
\label{eqn:Eckart_frame_base}
\end{equation} 
Here, $\mathcal{F}^{-1/2}$ represents a positively defined matrix, for which the following relation holds:
\begin{equation}
\mathcal{F}^{-1/2} \cdot \mathcal{F}^{-1/2} = \mathcal{F}^{-1},
\end{equation}
where $\mathcal{F}^{-1}$ is a positively defined inverse of the Gram matrix $\mathcal{F}$.

The reference components $c_i^\alpha$s are in the instantaneous Eckart frame given as:
\begin{equation}
\mathbf{c}_{\alpha} = \sum_{i=1}^3 c_i^{\alpha} \mathbf{f}_{i}.
\label{eqn:Eckart_frame_c}
\end{equation}
This  means  that  the  dynamics of  the  reference  configuration  is
governed by  the time evolution of  the positions of the  monomers. As
mentioned above,  the $c_i^\alpha$s  are in  general constant  and the
reference configuration  is rigid.  Besides, as  will be  shown below,
there is no  angular momentum with respect to  the internal coordinate
system in  the zero-th  order of displacement  of monomers  from their
reference positions~\cite{Wilson:1955}. Consequently,  the dynamics of
the reference configuration is nothing but the overall rotation of the
molecule,    which   is    described   by    the   angular    velocity
$\mathbf{\omegae}$.

The rotation  of  the  polymer is 
defined by the rotation of the base vectors of the Eckart frame:
\begin{equation}
\dot{\mathbf{f}}_{i} = \mathbf{\omegae} \times \mathbf{f}_{i}.
\label{eqn:Eckart_frame_rotation_base} 
\end{equation}
The combination of Eqs.~\ref{eqn:Eckart_frame_c} and~\ref{eqn:Eckart_frame_rotation_base} yields the following relation~\cite{Rhee:1997}:
\begin{equation}
\dot{\mathbf{c}}_{\alpha} = \mathbf{\omegae} \times \mathbf{c}_{\alpha}.
\end{equation}
It must be emphasized that there  are different ways how to attach the
initial  internal coordinate  system to  the reference  configuration.
Each of  these yields different  $c_i^\alpha$s and a  different Eckart
frame.  Nevertheless,  once the initial internal  coordinate system is
chosen,    the    Eckart    frame    is   defined    in    a    unique
way~\cite{Praprotnik:2005}.
The independence on the choice of  the {\em fixed} reference configuration
is illustrated in the Appendix,  where we report
results for three completely different reference configurations: (i) a
fixed  configuration taken  from a  frozen $T=0$  state; (ii)  a fixed
configuration adapted  to the  average molecular  shape found  at each
shear  rate  $\dot{\gamma}$ and  (iii)  a  mobile configuration  which
adapts  over  time  to  the  average  molecular  conformation  upon  a
pre-determined  ``averaging'' time  $\tau_w$.   We find  that the  two
fixed configurations give the same Eckart rotation frequency $\omegae$
while the third one consistently converges to the outcome of the fixed
references    for     $\tau_w\rightarrow    \infty$     while    for
$\tau_w\rightarrow 0$,  it provides the apparent  frequency $\omegal$
obtained from the standard approach.

\begin{figure}
\includegraphics[width=0.4\columnwidth]{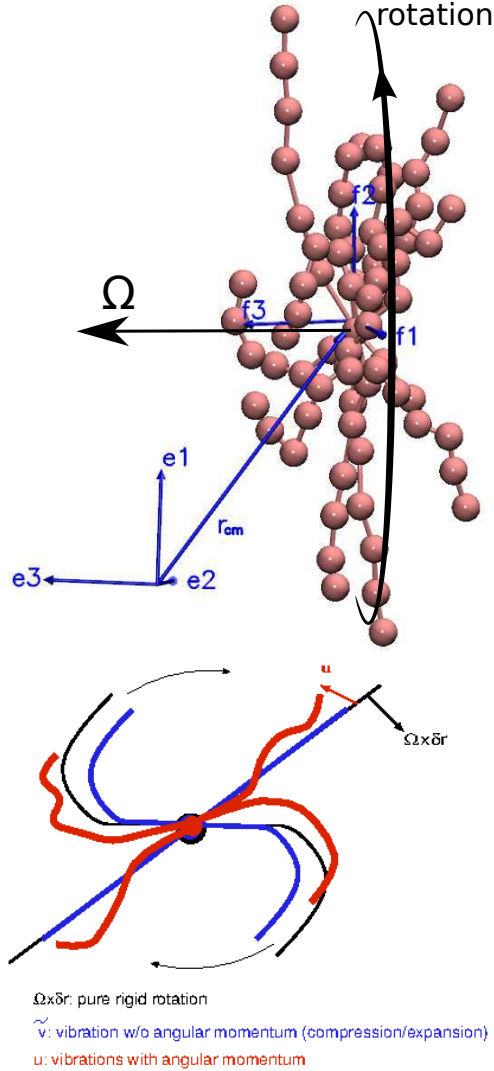}
\caption{(Top) A sketch of internal and laboratory frame. The unit base vectors $\mathbf{f_{1}}$, $\mathbf{f_{2}}$, and $\mathbf{f_{3}}$ span the internal coordinate system, i.e. the Eckart frame, which translates and rotates together with the molecule. The laboratory frame's base vectors are $\mathbf{e_{1}}$, $\mathbf{e_{2}}$, and $\mathbf{e_{3}}$. The arrows indicate the rotation of the molecule and its vibrations. (Bottom) The sketch
      gradually introduces the different types of
      displacements resolved by the Eckart formalism.
      The black line corresponds to pure rigid rotation
      (monomer velocity $\Omega\times \delta {\bf r}$) which
      does not introduce molecular deformation. 
      The blue line (velocity $\tilde{{\bf v}}$) introduces vibrations without angular momentum contribution (e.g. compression and expansion) and
      the red line introduces vibrations  with angular momentum  (fluctuations with velocity ${\bf u}$) which deform the molecule's shape (e.g. due to Brownian diffusion). Note that ${\bf u}\cdot \Omega \times \delta {\bf r}<0$ (Coriolis term).
      The different velocities are explained in the text (see e.g. Eqs. \ref{eqn:velocity decomposition} and \ref{eqn:vibrational velocity Eckart}).}
\label{fig:coordinate_system_sketch}
\end{figure}

The reference positions of every monomer in the laboratory frame are computed as:
\begin{equation}
\mathbf{d}_{\alpha} = \mathbf{r}_{cm} + \mathbf{c}_{\alpha},
\label{eqn:Eckart_frame_d}
\end{equation}
and their instantaneous displacement vectors are defined as 
\begin{equation}
{\bm \rho}_{\alpha} = \mathbf{r}_{\alpha} - \mathbf{d}_{\alpha}.
\label{eqn:Eckart_frame_rho}
\end{equation}
The unit base vectors of the Eckart frame $\mathbf{f_{1}}$, $\mathbf{f_{2}}$, and $\mathbf{f_{3}}$ satisfy the Eckart conditions\cite{Eckart:1935,Louck:1976}
\begin{equation}
\sum_\alpha m_\alpha{\bf c}_\alpha\times{\bm \rho}_\alpha=0,
\end{equation}
which state that there is no angular momentum with respect to the internal coordinate system in the zero-th order of displacements of the monomers from their equilibrium positions \cite{Wilson:1955}. The sketch of the Eckart frame for a star polymer is depicted in Fig.~\ref{fig:coordinate_system_sketch}. 

The angular velocity of the Eckart's coordinate system is given by: 
\begin{equation}
\mathbf{\omegae} = \mathbf{J'}^{-1} \cdot \sum_{\alpha=1}^{N} m_{\alpha} \mathbf{c}_{\alpha} \times \left( \dot{\mathbf{r}}_{\alpha} - \dot{\mathbf{r}}_{cm} \right).
\label{eqn:Eckart omega}
\end{equation}
The tensor $\mathbf{J'}$ is defined as
\begin{equation}
\mathbf{J'} = \sum_{\alpha=1}^{N} m_{\alpha} \lbrace \left[\left(\mathbf{r}_{\alpha} - \mathbf{r}_{cm}\right) \cdot \mathbf{c}_{\alpha}\right]\mathbf{I} - \left(\mathbf{r}_{\alpha} - \mathbf{r}_{cm}\right) \otimes \mathbf{c}_{\alpha} \rbrace .
\label{eqn:Eckart inertia}
\end{equation}
In the limit of a rigid molecule, Eq.~\ref{eqn:Eckart inertia} becomes Eq.~\ref{eqn:standard inertia tensor} and both definitions of angular velocity (given by Eqs.~\ref{eqn:standard eqn. omega} and ~\ref{eqn:Eckart omega}) are equivalent. 

The velocity of a given monomer $\alpha$ can be written as~\cite{Wilson:1955, Praprotnik:2005}:
\begin{equation}
\dot{\mathbf{r}}_{\alpha} = \dot{\mathbf{r}}_{cm} + \mathbf{\omegae} \times \left( \mathbf{r}_{\alpha} - \mathbf{r}_{cm} \right) + \Delta \mathbf{v}_{\alpha}.
\label{eqn:velocity decomposition}
\end{equation}
The first term on the right hand side of Eq.~\ref{eqn:velocity decomposition} represents the velocity of molecule's CoM, the second term is the contribution due to the rotation of the molecule, and the third one, i.e. $\Delta \mathbf{v}_{\alpha}$ due to molecular vibrations.
The latter can be expressed as~\cite{Rhee:1997}:
\begin{equation}
\Delta \mathbf{v}_{\alpha} = \mathbf{\tilde{v}}_{\alpha} + \mathbf{u}_{\alpha},
\label{eqn:vibrational velocity Eckart}
\end{equation}
where $\mathbf{u}_{\alpha}$ represents the angular motion part and $\mathbf{\tilde{v}}_{\alpha}$ (the same as in Eq.~\ref{eqn:velocity monomer standard}) the angular motion free part of the vibrational motion. Comparing the expressions in Eqs.~\ref{eqn:velocity monomer standard} and~\ref{eqn:velocity decomposition}, one derives the following equation~\cite{Rhee:1997}:
\begin{equation}
\mathbf{u}_{\alpha} = \left( {\bm \omegal} - \mathbf{\omegae} \right) \times \delta {\bf r}_{\alpha},
\label{eqn:diff velocity Eckart}
\end{equation}
with $\delta {\bf r}_{\alpha}\equiv {\bf r}_{\alpha}-{\bf r}_{cm}$. This means that $\mathbf{u}_{\alpha}$ is the part of $\alpha$-th monomer's vibrational motion, which is coupled with rotations if the angular velocity is calculated by the standard approach. It can be decoupled from rotations by using the Eckart frame formalism.

According to Eq.~\ref{eqn:velocity monomer standard},
the kinetic energy $T=\frac{1}{2} \sum_{\alpha} m_{\alpha}\dot{\mathbf{r}}_{\alpha}^{2}$ 
of any rotating molecule can be written as:
\begin{equation}
  T = \frac{1}{2} M \dot{\mathbf{r}}_{cm}^{2} + \frac{1}{2} {\bm \omegal}\cdot {\bf J}\cdot {\bm \omegal} + \frac{1}{2} \sum_{\alpha} m_{\alpha} \mathbf{\tilde{v}}_{\alpha}^{2},
\label{eqn:kinetic energy decomposition standard}
\end{equation}
where $M=\sum_{\alpha} m_{\alpha}$ the molecule's mass. These three terms
in the right hand side, are collected into,
\begin{equation}
T=T_{trans} + T_{rot}^{lab} + T_{vib}^{lab},
\end{equation}
with $T_{trans}$, $T_{rot}^{lab}$, and $T_{vib}^{lab}$
the translational, rotational, and vibrational contributions to the kinetic energy.

On the other hand, using the Eckart frame,
the velocity of each monomer is expressed by Eq.~\ref{eqn:velocity decomposition}
and the kinetic energy of a molecule is decomposed as~\cite{Rhee:1997}
\begin{equation}
  T =  \frac{1}{2} M \dot{\mathbf{r}}_{cm}^{2} + \frac{1}{2} {\bm \omegae} \cdot {\bf J} \cdot {\bm \omegae}
  + \frac{1}{2} \sum_{\alpha} m_{\alpha} \mathbf{\tilde{v}}_{\alpha}^{2} + \frac{1}{2} \sum_{\alpha} m_{\alpha} \mathbf{u}_{\alpha}^{2} + \sum_{\alpha} \mathbf{u}_{\alpha} \cdot \left(
  {\bm \omegae} \times \delta \mathbf{r}_{\alpha} \right).
\label{eqn:kinetic energy decomposition Eckart}
\end{equation}
One can now distinguish the following terms (in order of appearance
in the RHS of Eq. \ref{eqn:kinetic energy decomposition Eckart}):
\begin{equation}
T=T_{trans} + T_{rot}^{Eck} + T_{vib-non-ang}^{Eck} + T_{vib-ang}^{Eck} + T_{Cori}^{Eck}.
\end{equation}
Here,  $T_{rot}^{Eck}$   denotes  pure  rotational   contribution.  The
vibrational contribution  consists of  two parts: the  first, emerging
from the  angular free part of  the vibrational motion, is  denoted by
$T_{vib-non-ang}^{Eck}$,  and  the second,  i.e.  $T_{vib-ang}^{Eck}$,
representing the angular part of  vibrations. The last contribution is
the Coriolis coupling, which is denoted by $T_{Cori}^{Eck}$. Comparing
both kinetic energy expressions  (i.e. in Eqs.~\ref{eqn:kinetic energy
  decomposition  standard}  and~\ref{eqn:kinetic energy  decomposition
  Eckart}), we observe that the following relations hold
\begin{eqnarray}
  \label{eq:kin2}
  T_{vib}^{lab} &=& T_{vib-non-ang}^{Eck}, \\
  \label{eq:kin3}
  T_{rot}^{lab} &=& T_{rot}^{Eck} + T_{vib-ang}^{Eck} + T_{Cori}^{Eck}.
\end{eqnarray}
and obviously, the translational kinetic energy $T_{trans}$ is the same in both frames.
In order to alleviate the notation, we define the pure rotational energy
$T_{rot}^{Eck}$, the angular-momentum free vibrational energy
$T_{\tilde v}$, and the net vibrational angular momentum energy $T_{u}$ as,
\begin{eqnarray}
  T_{\Omega} &\equiv & \frac{1}{2} {\bm \omegae} \cdot {\bf J} \cdot {\bm \omegae}
  = T_{rot}^{Eck} \label{eq:trot},\\
  T_{\tilde v} &\equiv& \frac{1}{2} \sum_{\alpha} m_{\alpha} \mathbf{\tilde v}_{\alpha}^2
  =T_{vib}^{lab} = T_{vib-non-ang}^{Eck}, \\
  \label{eq:tu}
  T_{u} &\equiv&\frac{1}{2} \sum_{\alpha} m_{\alpha} \mathbf{u}_{\alpha}^{2} + \sum_{\alpha} \mathbf{u}_{\alpha} \cdot \left(
  {\bm \omegae} \times \delta \mathbf{r}_{\alpha} \right) =
  T_{vib-ang}^{Eck} + T_{Cori}^{Eck}.  
\end{eqnarray}

In the  next section, we resort  to the Eckart frame  formalism in the
analysis of the  rotational and vibrational behavior of star polymers  in solution and
melt. Differences with respect the laboratory frame will be highlighted.

\section{Results and discussion}

In this section, we present results from two different types of systems:
i) a single star polymer in solution (representing a dilute polymer suspension) and
ii) a melt of star polymers, with polymer volume fraction $\phi=0.2$ under isothermal conditions.
The molecular model of the star polymer is the same in both types of simulations. In the solution case, we consider two types of bonds between monomers (blobs):
harmonic bonds and finitely extensible non-linear elastic (FENE) bonds. In melt simulations, we use harmonic bonds
to build up the star molecule. 

We have so far introduced two frequencies (i.e. $\omegal$ and $\omegae$) describing
rotation in polymers. In what follows, we will introduce two additional frequencies and for the sake of clarity and reference,
we list them all in Table~\ref{tab:freq}.

\begin{table}
  \caption{The different frequencies mentioned in this work.}
  \begin{tabular}{c|cc}
    $\omegal$ & Apparent angular velocity & Eq. \ref{eqn:standard eqn. omega}\\
    $\omegae$& Eckart angular velocity & Eq. \ref{eqn:Eckart omega}\\
    $\omegar$& Monomer rotation frequency & Eq. \ref{eqn:tumbling_fit_function}\\
    $\omegat$& Breathing mode frequency & From Eq. \ref{eqn:tumbling cross-corr.}\\ 
\end{tabular} 
\label{tab:freq} 
\end{table}

To present the results in non-dimensionalized form, we use  the Weissenberg numbers $\wi$ and $\wi_{rot}$. The latter is based on
rotational   diffusion   time   of   the  star. This  is  defined  as
$\wi_{rot} = \dot{\gamma}\tau_{rot}$ where $\tau_{rot}$ is
the time for rotational diffusion $\tau_{rot}=R^2/D_r$ of the molecule
in equilibrium (see \cite{Sablic:2016, Sablic:2016:1} for details). The $\wi$, on the other hand, is based on the largest relaxation time ($\tau_{rel}$) of the molecule, i.e. $\wi = \gam \tau_{rel}$
It has to be said that the molecular rotational diffusion is the
slowest relaxation process for stars with harmonic bonds, while
for star molecules  with FENE bonds, the slowest  relaxation is the
process of arm disentanglement. The corresponding
relaxation times (rotational and arm-disentanglement) for simulations in solution and in melt
are given in Table~\ref{tab:relax. time}.

\begin{table}
  \caption{The rotational diffusion and arm-disentanglement
    relaxation times for our star model, with $12$ arms and $6$ monomers per arm.
    We define $\wi_{rot}=\dot{\gamma}\tau_{rot}$
    and  $\wi=\dot{\gamma}\tau_{rel}$ with
    $\tau_{rel}=\max\left[\tau_{rot},\tau_{dis}\right]$.} 
\begin{tabular}{ccc}
\centering
System & $\tau_{rot}$ & $\tau_{dis}$ \\
\hline
closed melt: $\gamma _{\parallel} = 1.0$, $\gamma _{\perp} = 1.0$ & $710 \pm 40$ & $390 \pm 10$\\
open melt: $\gamma _{\parallel} = 1.0$, $\gamma _{\perp} = 1.0$ & $700 \pm 40$ & $390 \pm 10$\\
solution harmonic bonds & $270 \pm 20$ & $180 \pm 20$\\ 
solution FENE bonds & $370 \pm 30$ & $950 \pm 90$\\
\end{tabular}
\label{tab:relax. time}
\end{table}

\subsection{Star polymer models}
The star polymer model is  taken from Ref.~\cite{Hijon_2010}.
We use the standard Lennard-Jones units, taking the monomer mass $m_{0}$, unit length $\sigma_0$
and energy $\epsilon_0$ as reference.
We consider stars with $f=12$ arms an $m=6$ beads per arm, with a total
of  $73$  monomers  (including   the  central  one).  Excluded  volume
interactions   of    monomers   are    modeled   by    the   repulsive
Weeks-Chandler-Anderson interaction ($\sigma =  2.415$ and $\epsilon =
1$). The bonds between adjacent monomers  $i$ and $j$ are modelled by
either harmonic springs or FENE bonds. In the case of
harmonic  bonds, with  a recovery  force  $-K(r_{ij}-r_{ij}^{eq})$,
the spring constant is $K =  20$ and the equilibrium distance  $r_{ij}^{eq} =
2.77$ (the  equilibrium distance between  the central monomer  and the
first monomer of an arm is larger $r_{ij}^{eq} = 3.9$). 
Finitely extensible bonds are modeled by the FENE potential \cite{Veldhorst:2015},
with a spring constant $K=20$ and maximum length of the bond $r_{max} = 1.5 \, r_{ij}^{eq}$.

\subsection{Melt simulations}
\begin{figure}
\includegraphics[scale=1.1]{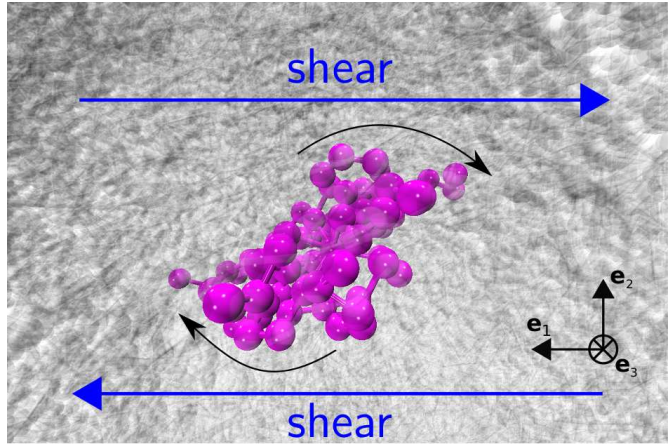}
\caption{Snapshot of the star-polymer melt under shear flow, drawn from the perspective of one polymer. The latter is depicted in purple and its surrounding polymers are colored in gray. The blue arrows correspond to the direction of the imposed shear, while the black arrows indicate the tank-treading rotation of the polymer. The coordinate unit vectors $\mathbf{e}_{1}$, $\mathbf{e}_{2}$, and $\mathbf{e}_{3}$ define the flow ($x_{1}$), the gradient ($x_{2}$), and the neutral ($x_{3}$) direction, respectively.}
\label{fig:melt_polymer_view}
\end{figure}

For the  melt case, we use  stars made of harmonic  bonds. Simulations
are carried out  at fixed temperature $T=4$  using molecular dynamics
with    a    dissipative    particle    dynamic    (DPD)    thermostat
\cite{DPD_Espanol,DPD_Soddemann}.  We  solve  systems  with  constant
volume (closed setup) and also open systems under constant normal load
(see Ref. for  details \cite{Sablic:2016}).  The simulation  box is of
size  $390 \times  117 \times  117$  and the  density of  the melt  in
equilibrium corresponds  to the  occupational factor  $\mathit{\Phi} =
0.2$, with about  $2000$ molecules. In the closed  periodic setup, the
shear flow  is imposed  by the SLLOD  algorithm implemented  with the
Lees-Edwards boundary conditions~\cite{LeesEdwards,  Sllod1, Sllod2}.  Constant  load
simulations  in an  open system,  are performed  using OBMD~\cite{hmd_prl06,Tools,Flekbus,DelgadoBuscalioni:2015},    which
permits to  impose an external  shear stress at  the open ends  of the
system. We shall  use the following coordinates: $x_{1}$
  refers  to the  flow  direction,  $x_{2}$ to  the  direction of  the
  velocity gradient  and $x_{3}$  to the  direction of  flow vorticity
  (sometimes  called neutral  direction).  The  DPD thermostat  used
here introduces friction  along the {\em normal}  and {\em tangential}
directions        of        any         pair        of        monomers
\cite{Sablic:2016,Hijon_2010,DPD_Espanol,DPD_Soddemann}   which   come
closer than the DPD-cutoff radius  $R_{DPD} = 2 \times 2^{1/6} \sigma$
(we  use a  Heaviside kernel  for the  DPD interaction).  The friction
coefficients  in normal  and tangential  directions equal $\gamma
_{\parallel}  = 1.0$  and $\gamma  _{\perp} =  1.0$. The  equations of
motion      are      integrated       by      the      Velocity-Verlet
algorithm~\cite{Tuckerman:2010} with the  integration step $0.01 \tau$
for  small  and  moderate  shear,  and $0.005  \tau$  for  high  shear
rates. A sketch of the star-polymer melt under shear flow
  from the perspective of one  of its constituent polymers is depicted
  in Fig.~\ref{fig:melt_polymer_view}.

\subsection{Star in solution}
We simulate a single star polymer in solution using Brownian hydrodynamics \cite{Jendrejack:2000, Jendrejack:2002}. The monomers (representing a coarse description of the molecule) interact via conservative forces (bonds and excluded volume interactions) and also via hydrodynamic interactions. The displacement of monomer $\alpha$ in direction $i$ over time $dt$ has the form $d r_i^{\alpha} = \dot{\gamma} x_2^\alpha \delta^{i,1} dt + \mu_{ij}^{\alpha\beta}(r_{\alpha\beta}) F_j^{\beta} dt + d\tilde{r}_i^{\alpha}$ where the first term indicates the shear flow (acting in $1$-direction) and the  mutual drag arises from the mobility tensor $\mu_{ij}^{\alpha\beta}$ which in present calculations consists on the Rotne-Prager-Yamakawa (RPY) approximation~\cite{Jendrejack:2000, Jendrejack:2002}. The Brownian displacement  $d\tilde{\bf{r}}$ satisfies a fluctuating dissipation (FD) relation for its covariance $\langle d \tilde{r}_i^{\alpha}  d \tilde{r}_j^{\beta}\rangle = 2 k_BT \mu_{ij}^{\alpha\beta} dt$ and to solve $d\tilde{\bf{r}}$ we use the Fixman's method~\cite{Jendrejack:2002}. The integration scheme is an explicit Euler scheme with the time step $dt=0.01 \tau$. 
In the present study, all simulations are run for $10000 \tau$.

\subsection{Monomer rotation dynamics}
\begin{figure}
\centering
\includegraphics[scale=0.44]{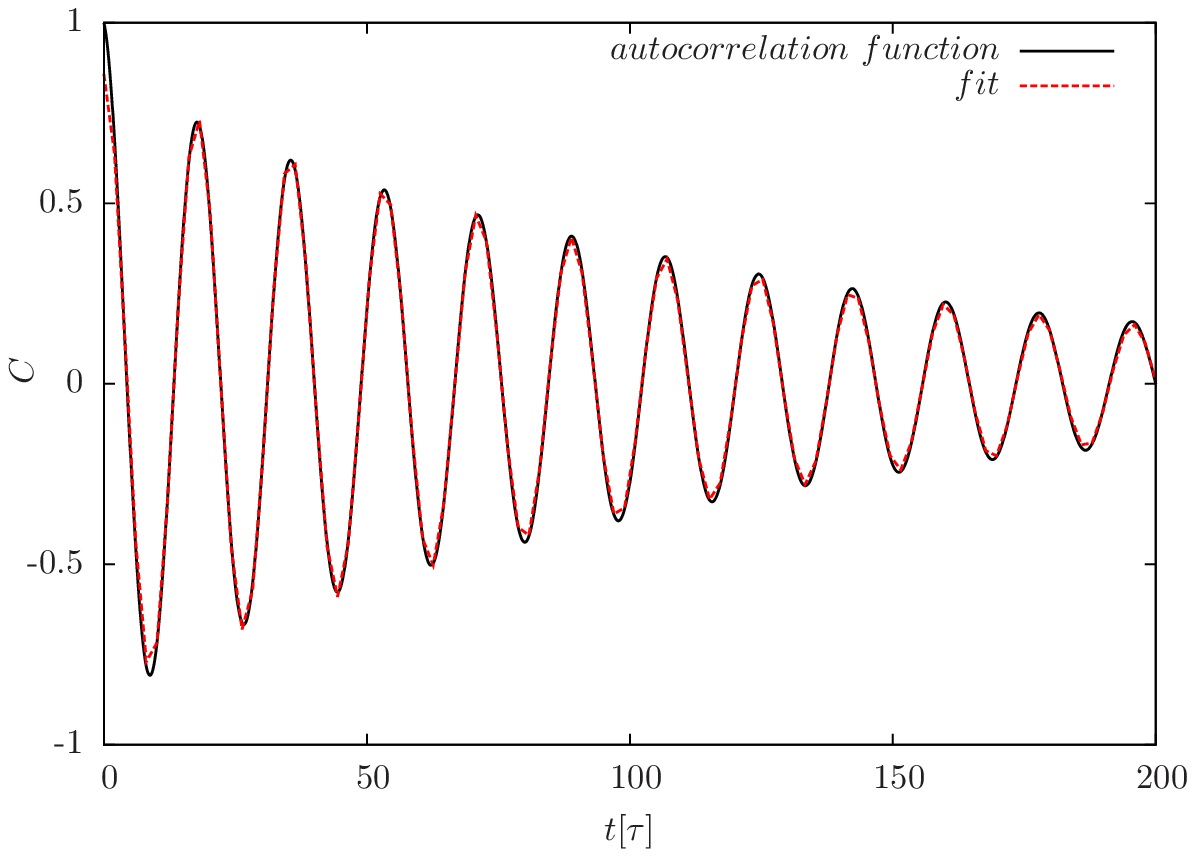}\\
\includegraphics[scale=0.5]{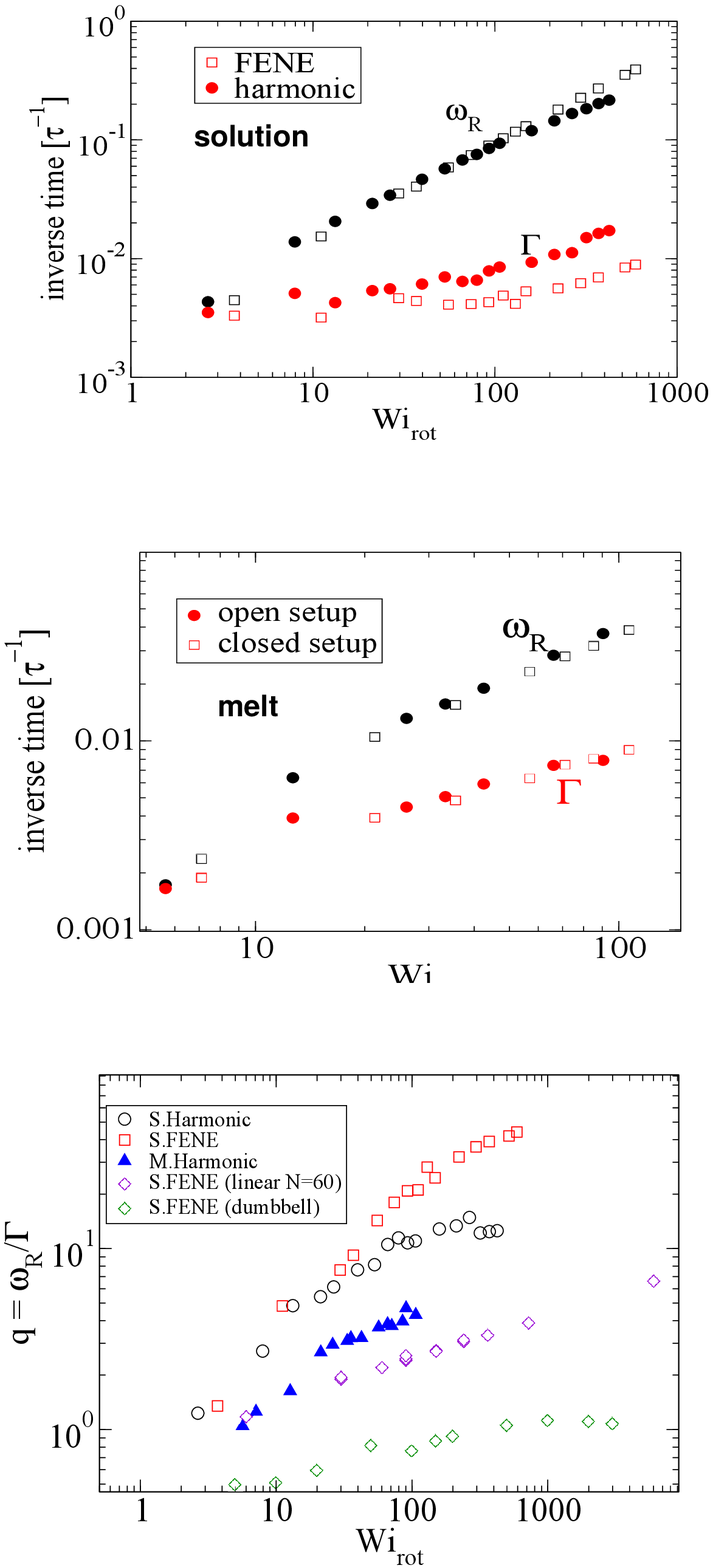}
\caption{(Top) Autocorrelation function of position of the final monomers of each polymer's arm in the gradient direction, fitted by Eq.~\ref{eqn:tumbling_fit_function} with parameters: $A = 0.93$, $\omegar = 0.35$, $\Gamma = 0.0084$, and $\Psi = 0.0025$. (Middle panels) The tank-treading frequency $\omegar$ and decorrelation rate $\Gamma$ obtained from the fits (stars in solution and in melt). (Bottom) The quality factor $q=\omegar/\Gamma$ for the dynamics of monomer rotations,
    comparing our 12-6 stars with linear FENE chains ($N=60$) (with excluded volume interactions) and dumbbells, from Ref.\cite{Usabiaga:2011}.}
\label{fig:omega_d_fit}
\end{figure}

To provide direct connection  with  the
monomer  dynamics,  we calculate the  angular  velocity  of
rotation of  molecules from  the autocorrelation function of
the gradient-direction coordinate  of the last  monomer of  every arm of  the star,
relative to  the CoM, $({\bf r}_{\alpha}-{\bf r}_{cm})\cdot \hat{\bf x}_2$. This signal is similar to an
underdamped  oscillator  (Fig.~\ref{fig:omega_d_fit})   which  can  be
fitted with the following function~\cite{Usabiaga:2011}:
\begin{equation}
C\left(t\right) = A^{2} \cos\left(\omegar t + \psi\right)\exp\left(-\Gamma t\right),
\label{eqn:tumbling_fit_function}
\end{equation}
where the damping rate $\Gamma$ represents the decorrelation rate,
$\omegar$ the rotation frequency, and $\psi$ a phase constant.
Two issues are  noticeable from this graph: first, the decorrelation rate $\Gamma$ only
starts to significantly increase above $\wi_{rot}>50$. Second, as $\wi$ increases,
the quality factor $q=\omegar/\Gamma$ becomes quite large,
in particular, compared with what happens in linear polymers under shear \cite{Usabiaga:2011}
(which tumble by compressing, like in a tube).
Figure \ref{fig:omega_d_fit} (bottom panel) compares the quality factor $q$ for star polymers in solution (S) and melt (M)
(with either FENE or harmonic bonds) and that measured in Ref.\cite{Usabiaga:2011} for FENE linear chains with $N=60$ and dumbbells.
In the case of star molecules, the arms rotate almost like in a ``wheel'' and
a monomer turns around several times ($q$) before decorrelating its initial ``rigid-body'' position.
At large shear rates, the differences in values of $q$ are significant [see Fig. \ref{fig:omega_d_fit} (bottom)].
The quality factor is significantly smaller in melts,
indicating the hindrance arising from steric interaction amongst close-by molecules.
In what follows, we will compare $\omegar$ with $\omegal$ and $\omegae$
and discuss the origin of the decorrelation $\Gamma$, according to the Eckart analysis.

\subsection{Kinetic energies}

The kinetic  energy balance is illustrated  in Table~\ref{tab:kin} for
star molecules  in solution (having  harmonic or FENE bonds)  and some
values of the shear rate. Displacements describing pure rotations have
kinetic  energy  $T_{\Omega}$  but, coherent  (collective)  vibrations
without angular momentum contribute with the largest energy $T_{\tilde
  v}$.   These  are related  to  overall  shape deformations  (and  in
particular,   compression/expansion   does   not   introduce   angular
momentum). Other  type of molecular  deformations (affine or  not) are
collected  in  the  velocity  ${\bf u}_{\alpha}$  which  does  provide
angular momentum  (see Eq.  \ref{eqn:diff velocity Eckart})  and feeds
the (negative) kinetic  energy contribution $T_u=T_{vib-non-ang}+T_{Cori}$
(see Table \ref{tab:kin} and  Eq. \ref{eq:kin2}). Equation \ref{eq:tu}
confirms that this energy can only be negative because of the Coriolis
term. So, in  average, ${\bf u}\cdot \left(\Omega\times \delta
  {\bf  r}\right)<0$;  in  other  words ${\bf  u}$  contributes  in  opposite
  direction to  the pure  rotation velocity $\Omega\times  \delta {\bf
    r}$.   Note that  $|T_{u}|$  is subtracted  to  the pure  rotation
  energy $T_{\Omega}$  to yield the  total rotation kinetic  energy in
  the  lab  frame  $T_{rot}^{lab}$ (see Eq. \ref{eq:kin3}).   To  clarify  matters,  a  sketch
  illustrating  the  different  types  of displacements  is  drawn  in
  Fig.  \ref{fig:coordinate_system_sketch}  (bottom  panel).  In  what
follows, we analyze these kinetic energies separately.

\begin{table*}
\small
  \caption{\ Kinetic energy balance for solution of star polymers with harmonic and FENE bonds. The error bar of the reported values is approximately $5 \%$.}
  \label{tab:kin}
  \begin{tabular*}{\textwidth}{@{\extracolsep{\fill}}lllll|lllll}
    \hline
    \multicolumn{5}{c|}{solution harmonic bonds} & \multicolumn{5}{c}{solution FENE bonds} \\
    \hline
    $W_{i}$ & $T$ & $T_{\Omega}$ & $T_{\tilde{v}}$ & $T_{u}$ & $W_{i}$ & $T$ & $T_{\Omega}$ & $T_{\tilde{v}}$ & $T_{u}$\\
    \hline
    13.25 & 1102 & 412 & 1028 & -338 & 9.5 & 875 & 303 & 794 & -222\\
    53 & 1117 & 517 & 1042 & -442 & 95 & 876 & 365 & 793 & -282\\
    106	& 1135 & 464 & 1058 & -387 & 570 & 912 & 498 & 808 & -394\\
 	424	& 1363 & 1189 & 1275 & -1101 & 1520 & 1086 & 763 & 919 & -596\\
    \hline
  \end{tabular*}
\end{table*}

\subsection{Pure rotation: tank-treading}

\begin{figure}
\includegraphics[width=0.43\columnwidth]{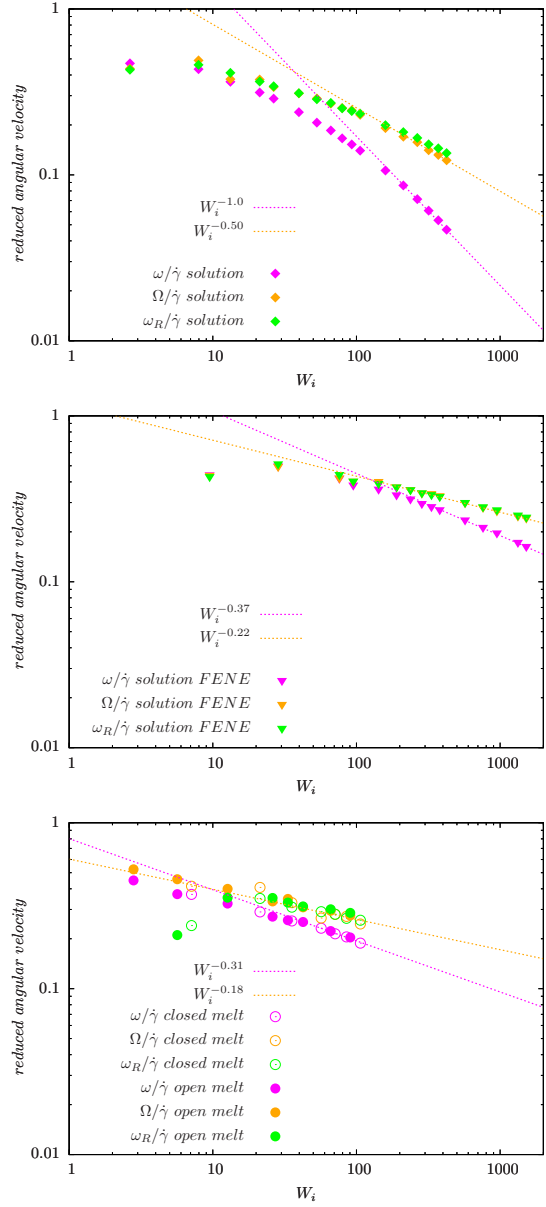}
\caption{Comparison of the angular  velocity computation by the Eckart
  frame formalism  ($\omegae/\dot{\gamma}$ - colored  in orange),
  by  the standard  approach ($\omegal/\dot{\gamma}$  - colored  in
  magenta), and by the autocorrelation function of the position of the
  final  monomers in  every polymer's  arm in  the gradient  direction
  ($\omegar/\dot{\gamma}$ -  colored in green). We  study rotations
  in the solution  of star polymers with Hookean (Top) and FENE
  bonds (Middle)  and in the melt of star  polymers with Hookean
  springs (Bottom).  In all three systems,  the angular velocity
  obtained  by the  Eckart  frame  formalism is  higher  than the  one
  calculated by the  standard approach. In all cases, $\omegar$   matches  well   with
  $\omegae$ while the difference $\omegal-\omegar$
  is larger in the  Hookean spring solution case, followed  by the  solution of
  molecules with the  FENE bonds and the melt. The  reasons for these facts are explained   in   the  text.}
\label{fig:rotation_Eckart_vs_standard}
\end{figure}

Figure~\ref{fig:rotation_Eckart_vs_standard} compares  the results for
the apparent  angular velocity $\omegal$, the  Eckart angular velocity
$\omegae$  and  the  frequency  of monomers  rotation  about  the  CoM
$\omegar$.  In all considered cases (polymers with either harmonic or FENE bonds
in solution, and the melt case), we find that $\omegae=\omegar$ within error bars while $\omegae>\omegal$.
Whenever vibrational angular momentum is present, the apparent angular
velocity  $\omegal$ does  not correctly represent molecular  rotation
\cite{Rhee:1997}.  The  difference between $\omegal$ and  $\omegae$ is
larger    for   stars    with   Hookean-bonds    in   solution    (see
Fig.  \ref{fig:rotation_Eckart_vs_standard}).
From Eq. \ref{eqn:diff velocity Eckart}, this simply  indicates
that vibrational angular momentum ($\mathbf{u}_{\alpha}$) has a larger
contribution if the molecule is softer (harmonic
versus FENE bonds) or has more free space to deform (as in the case of
solution  compared  to melt). 

Stars with harmonic bonds in solution seems to reach the scaling
$\omegal/\dot{\gamma} \sim \wi^{-1}$ (i.e. $\omegal \rightarrow  \mathrm{cte}$)
as the shear rate is increased (although, in fact, at very large $\dot{\gamma}$, $\omega$
decreases). This apparent scaling
was attributed in Ref.  \cite{Ripoll:2006} (and subsequent citations)
to an universal limiting trend for tank-treading rotation of star polymers. However, although the
apparent  angular  velocity  $\omega$ reaches a maximum value,
the  tank-treading     frequency  $\omegar$, keeps increasing with $\dot{\gamma}$, like
$\omegar\sim \wi^{\alpha}$ with $\alpha=0.5\pm 0.02$.
This is shown in Fig.~\ref{fig:rotation_Eckart_vs_standard})
where one can see that $\omega$ and $\omegar$ differ significantly.

Finally, in melts, (bottom  panel  of
  Fig.~\ref{fig:rotation_Eckart_vs_standard}) we   observe  that  the
  molecular rotational frequencies are similar in
  the open and closed environments. This is in agreement with
  our previous studies  (Ref.~\cite{Sablic:2016, Sablic:2016:1})
  and indicates that the rheological differences measured
  in open and closed environments are of thermodynamic origin
  (density decreases when an open polymer enclosure is sheared).

\subsection{Vibrational angular momentum and decoherence of rotational motion}

Following  this  line,  Eq.  \ref{eq:kin3} indicates  that  the  total
kinetic energy coming from displacements  with angular momentum can be
decomposed in a pure rotational part $T_{\Omega}$ and contributions
from  vibrational  angular  momentum.   It  is  noted  that
$T_{\Omega}$ contains   contributions   from  collective displacements
and also from fluctuations. Equation \ref{eq:trot} indicates that
\begin{equation}
T_{\Omega} = \frac{N}{2} \omegae_3^2 \left(G_{11}+G_{22}\right) + \tilde{T}_{rot},
\end{equation}
where $\tilde{T}_{rot}$ introduces a significant contribution
from the covariances involving zero-average components of the rotational frequency $\boldsymbol{\Omega}$,
like $\tilde{T}_{rot} =\langle N \omegae_1^{2} G_{22}\rangle + ...$.
Here, $G_{ii}$ represents the diagonal gyration tensor component in the $i$-th direction.

\begin{figure}
  \includegraphics[width=0.5\columnwidth]{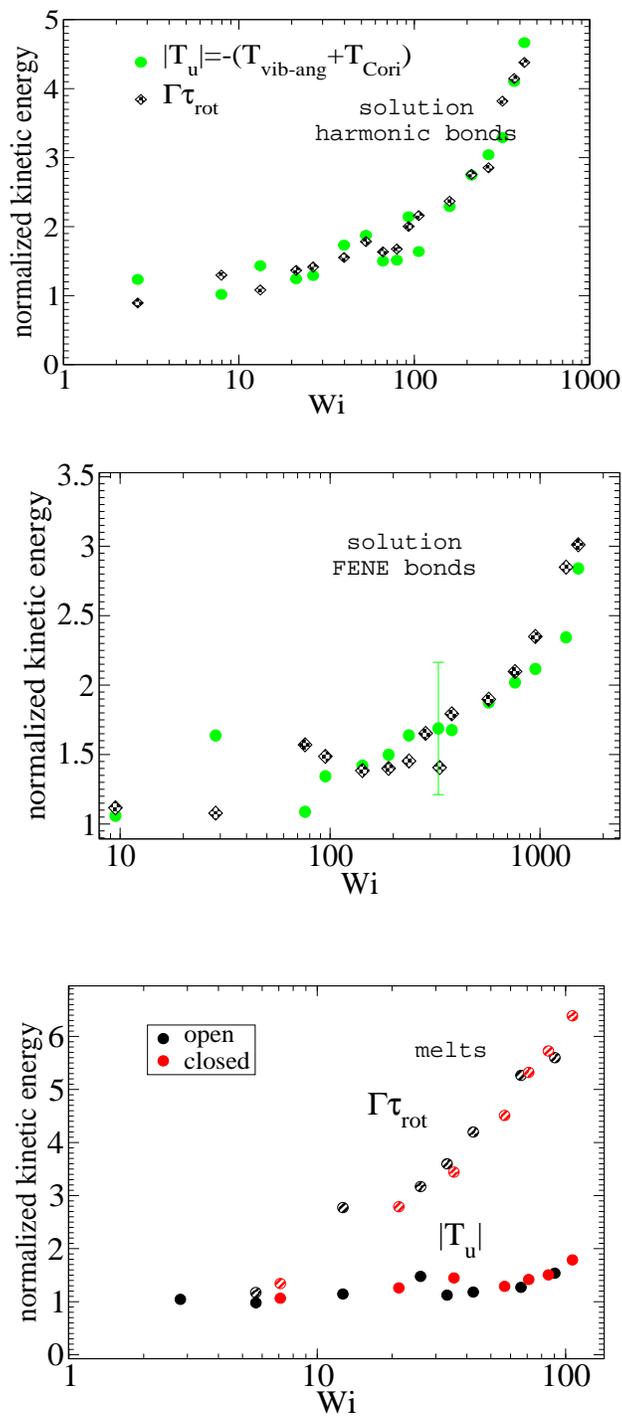}
  \caption{The absolute value of kinetic energy related
    to vibrational non-angular momentum $|T_{u}|$
    compared with the rate of decorrelation ($\Gamma$)
    of the monomer pure rotation around the molecule center
    (see Fig. \ref{fig:omega_d_fit}). Both quantities are normalized
    with their values at zero shear rate. Top and middle panel,
    results for solution and bottom panel, for the melt.}
\label{fig:dis}
\end{figure}

The energy of vibrations with angular
momentum  corresponds  to deformations of the  arms away from pure  rigid body
rotation (see Eq. \ref{eqn:diff velocity Eckart}).
In solution, these motions arise from Brownian diffusion
so we expect that the kinetic  energy  $|T_u|$
is proportional  to  $\Gamma  D_{arm}$  where
$D_{arm}$ is the diffusion coefficient of  the center of mass of
one star's arm (which is independent on the shear rate).
The scaling this hypothesis  predicts is   validated    in
Fig. \ref{fig:dis} where $|T_u|$
(normalized with its value at zero shear rate)
is compared with  $\Gamma\tau_{rot}$
for  increasing  Weissenberg  number.  Results  for
different types  of star  polymers (harmonic  and FENE  bonds) confirm
that both magnitudes are proportional and indicate that our intuition
contains  physical insight. In melts, however, both quantities
differ significantly (see Fig. \ref{fig:dis} bottom panel)
indicating that, in this case, molecular deformations are also
determined by other (non-Brownian) mechanisms, like inter-molecular collisions.

\subsection{Vibrations without angular momentum and breathing mode}

\begin{figure*}
  \includegraphics[scale=0.55, angle=-90]{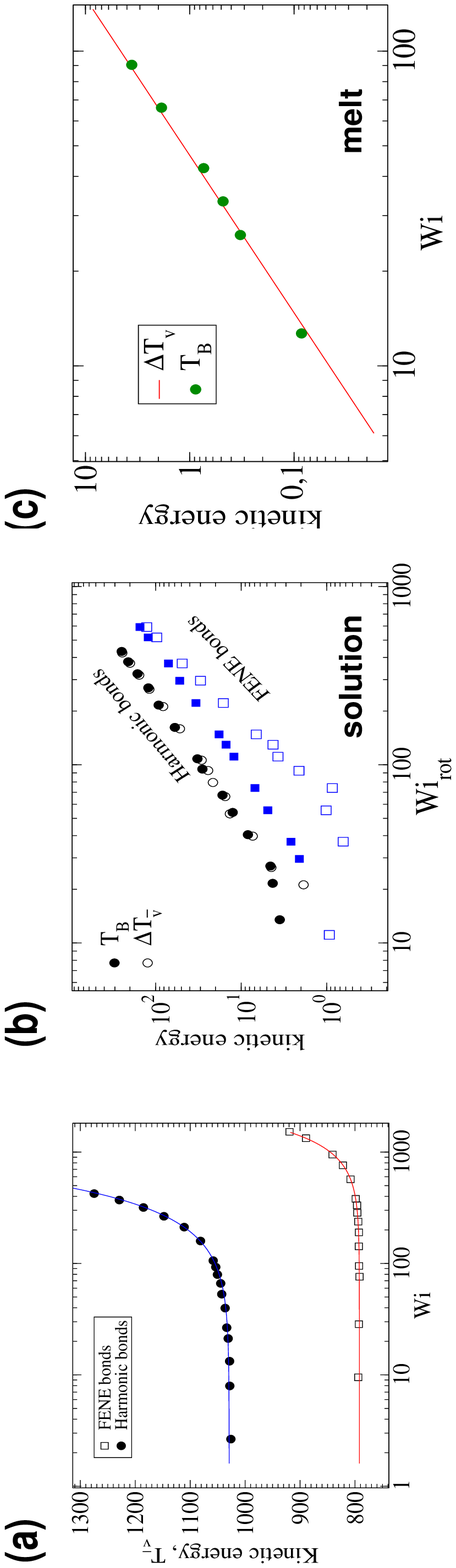}
  \caption{(a) The  angular momentum free vibrational kinetic $T_{\tilde v}$
    (symbols) and the fit $T_{\tilde v}(\wi) =T_{\tilde v}(0)+ a \wi^{\beta}$
    with $\Delta T_{\tilde v} = a \wi^{\beta}$ the coherent part and
    $T_{\tilde v}(0)$ the thermal contribution (results for star in solution). (b) The coherent contribution
    $\Delta T_{\tilde v}$  is compared with the breathing mode energy estimated as
    $T_B=\frac{N}{2} \omegat^2 \mathrm{Std}[G_{1}]$,  where $G_{1}$ is
    the principal eigenvalue of the gyration tensor and $\omegat$ is the breathing frequency,
    reported in Ref. \cite{Sablic:2016:1} (results for solution). (c) The same as (b) but
    for the melt case, and $T_B=0.45\,\frac{N}{2} \omegat^2 \mathrm{Std}[G_{1}]$. }
\label{fig:vib}
\end{figure*}

As stated (see Table \ref{tab:kin}), vibrations without angular momentum
$T_{\tilde v}$ have the largest contribution to the kinetic energy of the star molecule.
This kinetic energy has also a thermal and a coherent contribution.
The thermal energy includes the fluctuations in bond length,
whose average kinetic energy scales like $N_{sp} K \langle \delta^2\rangle $,
with $\delta=r_{\alpha,\beta}-r_{eq}$ the bond length,
$N_{sp}=72$ the number of springs in our star molecules
and $K$ their spring constant. Excluded volume forces are also central forces
${\bf F}_{\alpha,\beta}\propto {\hat {\bf r}}_{\alpha,\beta}$ (${\hat {\bf r}}_{\alpha,\beta}$ being the unit distance vector between monomers $\alpha$ and $\beta$)
so in absence of hydrodynamic interactions they strictly
do not contribute to the total angular momentum. It is noted that
hydrodynamics spreads over internal forces,
and contributes
to the angular momentum, with monomer displacements
$d{\bf r}_{\alpha}= \boldsymbol{\mu}_{\alpha\beta}{\bf F}_{\beta}\,dt$,
where $\boldsymbol{\mu_{\alpha\beta}}$ is the mobility tensor.
However, as shown in Ref. \cite{Sablic:2016:1},
the major source of angular momentum comes out from the mean flow.
We assume that  thermal contribution to $T_{\tilde v}$
is independent on the shear rate. The remaining contribution
to $T_{\tilde v}$ is assumed to be associated to overall deformations
of the molecular shape and should  increase with $\dot{\gamma}$.
This separation between thermal and coherent vibrations is clearly
revealed in the fit
$T_{\tilde v}(\wi)= T_{\tilde v}(0) + \Delta T_{\tilde v}(\wi)$,
which is shown in Fig. \ref{fig:vib}, with
$T_{\tilde v}(0) = 1029\pm 5$ and
$\Delta T_{\tilde{v}}(\wi) = 0.021\,\wi^{1.54}$ for harmonic springs and
while $T_{\tilde v}(0) = 792\pm 2$ and $\Delta T_{\tilde v}(\wi) = 2.32\times 10^{-5} \wi^{2.12}$
for FENE bonds (both in solution). In the case of melts we find
$T_{\tilde v}(0)=426$ and $\Delta T_{\tilde v}(\wi) =4.60\times10^{-4} \wi^{2.00}$.

We expect that the coherent part of the vibrational energy $\Delta    T_{\tilde v}(\wi)$
comes out from a collective ``oscillation'' of the molecule shape.
Such type of collective vibration was discussed in a previous work
on star polymers \cite{Sablic:2016:1}, and was referred to as ``breathing mode''.
The dynamics of the breathing mode is revealed in
the time correlation of the components of the
gyration tensor ($G_{ij}$), given by~\cite{tumbling, Chen:2013, Sablic:2016:1},
\begin{equation}
C_{ij}\left(t\right) = \frac{\langle \delta G_{ii}\left(t_{0}\right) \delta G_{jj}\left(t_{0} + t\right)\rangle}{\sqrt{\langle \delta G_{ii}^{j}\left(t_{0}\right)\rangle \langle \delta G_{jj}^{2}\left(t_{0}\right)\rangle}}.
\label{eqn:tumbling cross-corr.}
\end{equation}
where $\delta G_{ii} = G_{ii} - \langle G_{ii} \rangle$.
These are damped oscillatory signals with a characteristic frequency $\omegat$.
In previous works \cite{Chen:2013, Sablic:2016:1}, the cross-correlation $C_{12}$ has been used
to extract the ``tumbling'' time $\tau _{t}$
(as twice the difference between first maximum and first minimum).
We define $\omegat= 2\pi/\tau_t$.
As explained in Ref. \cite{Sablic:2016:1}, these type of dynamics
have been called ``tumbling'' in linear and ring chains, while
the word ``breathing'' is more appropriate to describe the star
overall shape oscillation, while they perform tank-treading.
The energy of ``breathing'' can be estimated from the
largest fluctuation in the gyration tensor, taken
from the standard deviation of the principal eigenvalue of the gyration tensor ${\bf G}$,
i.e. $\mathrm{Std}[G_{1}]= \langle(G_{1}-\langle G_{1}\rangle)^2 \rangle^{1/2}$.
A rough estimation of the breathing kinetic energy is then,
$T_{B} \equiv \frac{N}{2} \omegat^2 \mathrm{Std}[G_{1}]$, and it is compared
with $\Delta T_{\tilde v}$ in Fig. \ref{fig:vib}. In passing, we note that
a quite similar outcome is obtained by $T_B\propto \omegat^2 \mathrm{Std}[V^{2/3}]$
which is based on fluctuations (expansion/contraction) of the overall molecular volume $V=\prod_{\alpha} G_{\alpha\alpha}^{1/2}$. Interestingly, we find an excellent agreement (even quantitative) in all cases involving stars with harmonic bonds (solution and melt). However, in the FENE case, the values of $T_B$ and $\Delta T_{\tilde v}$
differ at small and moderate shear rate, and become similar as $\wi$ increases.
For moderate and small $\wi$ we find $\Delta T_{\tilde v} <T _B$, indicating
that the stronger excluded volume forces in FENE bonds (arm elongations are confined to a fixed value)
tend to reduce  collective vibrations (breathing) of the star molecules.

\subsection{Intrinsic viscosity}
\begin{figure*}
  \includegraphics[width=\columnwidth]{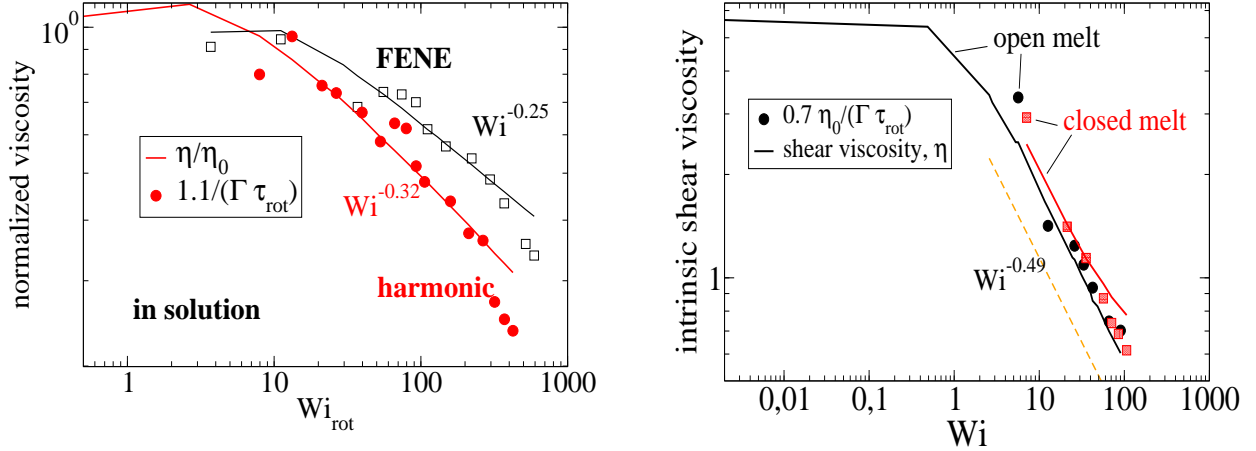}
  \caption{The intrinsic shear viscosity $\eta$ is compared with the normalized decorrelation
    rate of  the arms (center-to-end) distance  $\Gamma \tau_{rot}$.
    The  left panel corresponds to stars in dilute solution
    (here, we normalize with the viscosity  at zero  shear rate  $\eta_0$)
    and the right panel to stars in melt (polymer  volume fraction $0.2$).
    In  solution, the shear stress scales like
    $\sigma_{12}  = c  \dot{\gamma}/\Gamma$  with $c=15$  for harmonic  and
    $c=7$ for FENE bonds.}
\label{fig:eta_vs_Wi}
\end{figure*}

One of the major tasks of polymer physics is to relate individual chain dynamics with macroscopic rheological properties.
We make such an exercise in this section, taking the shear viscosity as our target macroscopic quantity.
In a previous work, we analyzed in some detail the rheology of these stars in melt \cite{Sablic:2016:1} and reported in particular
its shear viscosity under shear. Here, we calculate
the contribution to the shear viscosity of star polymers in solution from their contribution to the
virial part of the shear stress tensor \cite{Doyle:2005},
\begin{equation}
{\bm \sigma} =\rho_P  \Bigg \langle \sum_{\alpha = 1}^{N} \left(\mathbf{F}^{nb}_{\alpha} + \mathbf{F}^{b}_{\alpha}\right) \otimes \left(\mathbf{r}_{\alpha} - \mathbf{r}_{cm}\right) \Bigg\rangle.
\end{equation}
Here, $\mathbf{F}^{nb}_{\alpha}$ represents the force on the $\alpha$-th monomer, originating from the non-bonded interactions (i.e. the Weeks-Chandler-Anderson interaction), and $\mathbf{F}^{b}_{\alpha}$ are the forces of the bonds (i.e. either harmonic or FENE). The polymer contribution to the  stress tensor is proportional to $\rho_P$, the number density of polymer molecules,
and the polymer contribution to the shear viscosity is ~\cite{Doyle:2005}
\begin{equation}
\eta = - \frac{\sigma_{12}}{\dot{\gamma}}.
\end{equation}
Using  the  Carreau  fit~\cite{Yasuda:2006, Aho:2011},  we estimate  the  zero-shear  rate  viscosity
$\eta_0$ and present the normalized viscosity $\eta/\eta_0$. We note that $\eta_0$ is  about  1.8  times   larger  in  the  case  of  the
harmonic-bond model compared with the FENE bonds. As the shear rate is
increased, we find  shear thinning $\eta \sim  \wi^{-\beta}$ with shear
thinning exponents  $\beta=0.25$ for  FENE bonds and  $\beta=0.32$ for
harmonic bonds. These values are  somewhat smaller than those found in
melt, $\beta=0.49$ (see  Fig.\ref{fig:eta_vs_Wi}).
Viscous dissipation
is related to decorrelation times and in fact, the intrinsic viscosity
can be expressed as an sum  of relaxation times \cite{Doi1994}. For an
isolated star in dilute solution,  one expects that the main mechanism
for dissipation  comes from  the decorrelation  in arm  lengths, which
takes    place   at    an    average   rate    $\Gamma$   (see    Fig.
\ref{fig:omega_d_fit}).  Thus, as a first estimate, we seek a relation
of the  form $\eta  \propto \Gamma^{-1}$.   Figure \ref{fig:eta_vs_Wi}
shows that such  relation holds relatively well, both  in solution and
melts.   For instance,  in solution  we see  that the  softer harmonic
bonds  leads  to  faster  decorrelation rates  and  smaller  intrinsic
viscosity, compared with the more  rigid FENE chains.  As we indicated
in Fig.   \ref{fig:dis}, we found  that, in solution,  $\Gamma$ scales
like the kinetic energy $|T_u|$ and consistently, $|T_u|$ is larger in the
case of  harmonic bonds compared  with FENE-stars. In  melts, however,
one expects that  the departure from rigid-body  rotation (measured by
the  velocity $u$  and  its  kinetic energy  $T_u$)  arises also  from
inter-molecular  collisions (and  not only  from Brownian  diffusion).
This  is  revealed in  the  different  trends  followed by  $T_u$  and
$\Gamma$ in  melts: unlike  what it is  observed in
solution, $T_u$ and $\Gamma$ do  not correlate (see Fig. \ref{fig:dis}
botom).

\section{Conclusions}

The main purpose of this work is to show that the Eckart formalism
can be used to unveil the complex dynamics of  soft molecules in flow.
The application of the Eckart formalism to the  dynamics of star molecules in shear flow
permitted us to warn about  the incorrect interpretation of the rotation dynamics of soft molecules (polymers)
based  on  a  standard  (lab frame)  analysis.
In particular, the  {\em apparent} angular  velocity $\omegal$ resulting
from  such analysis has not a clear dynamical interpretation
(it is not the rotation frequency of  the molecule).
We have  shown that  the Eckart  co-rotating frame
correctly extracts the different types  of motions in the rotating and
vibrating molecule: pure rotation,  vibration with no-angular momentum
and vibrational angular  momentum. Star  molecules in  shear  flow  perform
a tank-treading  motion \cite{Ripoll:2006}  whereby monomers rotate  around the
center of the molecule, but for a given fixed shear rate, the molecule
keeps  a {\em  roughly fixed}  ellipsoidal overall  shape (more
precisely, they do not  tumble).  At  large  shear  rates,  the  molecule
performs  another  collective  motion,  which  we  called  ``breathing
mode'' \cite{Sablic:2016:1}, whereby the gyration tensor of the molecule oscillates in time
with a characteristic frequency $\omegat$.  We have shown that each of
these dynamics is associated with  a different type of displacement in
the Eckart frame.  The pure  rotational component of the Eckart frame,
with a  frequency $\omegae$, describes the  tank-treading frequency of
the star  $\omegar$.  By extracting  the thermal (incoherent)  part of
the kinetic  energy of vibrations without  angular momentum component,
we find that  the kinetic energy of the breathing  mode coincides with
the energy of ``breathing'' vibrations.  Finally, in solution,
we find that the decorrelation  of the end-to-end arm distance, driven
by Brownian diffusion at a  rate $\Gamma$, correlates with the kinetic
energy associated to vibrations  (or more properly, fluctuations) with
angular momentum,  $T_u$. In melt,  such correlation is  not observed,
and  it seems  that the energy $|T_u|$
of molecular  deformations is mainly determined  by
intermolecular collisions (and thus density dominated).

In this work, we just consider star polymers with $f = 12$ arms and $m = 6$ monomers per arm.
According to a recent analysis  \cite{Chremos:2015}, star molecules become chain-alike for $f<6$ so our stars are within the ``colloidal-alike'' regime. But, what would be the dynamics of more massive stars? While this question is open to future works, we have good reasons to believe that they will be quite similar to that found for $f=12, m=6$. In fact, several computational works for star polymers in dilute \cite{Ripoll:2006} and semidilute  \cite{Chen:2013} conditions, covered a relative large range of values of $f\leq 50 $ and $m< 50$ and (by defining the proper Weissenberg number) they found that all data for $\omega$ collapse in a master curve, indicating that the length of the arms or the functionality was not essentially changing the polymer dynamics. The dynamics would surely change in case of a semidilute solution (or melt) if the stars have {\em very} long arms ($m>100$), because entanglements should play a mayor role in distorting their rotation dynamics. However, we emphasize that the Eckart framework would be still applicable in such regime and provide valuable dynamic information. 

It also has to be noted that the present analysis
 can be complementary to the more detailed normal mode analysis of vibrations, within the framework of the theory of molecular vibrations~\cite{Eckart:1935}. In the latter, each internally rotating part of the molecule would require the introduction of additional internal coordinate systems inside the translating and rotating Eckart frame~\cite{Praprotnik:2005, Praprotnik:2005:1, Praprotnik:2005:2, Praprotnik:2005:3, Howard:1937:1, Howard:1937:2, Kirtman:1962, Kirtman:1964, Kirtman:1968}. Presently, we leave this discussion for the future work, since the main aim of this paper is the separation of rotations from vibrations, or the consequences such decomposition brings up in the interpretation of molecular rotations.
The objective of this work is to show that the Eckart frame, successfully and routinely used to describe
Raman spectra of small molecules, is also a robust and useful tool to investigate
the complex dynamics of soft, semiflexible macromolecules.

\appendix
\section*{Appendix: The Eckart reference configuration}
\begin{figure}
\includegraphics[scale=0.7]{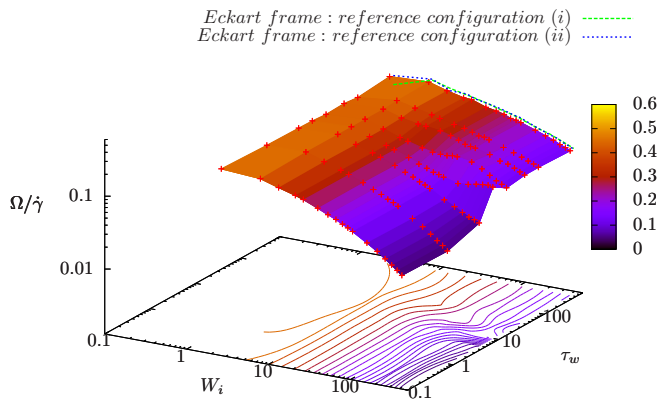}
\caption{Three definitions of reference configuration to calculate the angular velocity by Eckart frame formalism for star polymers in solution: \textbf{(i)} The reference configuration is the equilibrium configuration at temperature $0$~K (green line). \textbf{(ii)} We obtain the reference configuration at every shear rate separately by the Monte Carlo simulation so that it matches the average steady state shape of a polymer (i.e. gyration tensor) at that particular shear rate (blue line). \textbf{(iii)} The reference configuration is taken to be an instantaneous one, but in this way defined $c_{i}^{\alpha}$s are used in computation of the angular velocity only for the following $\tau _{w}$ in time. Afterwards, the reference configuration is replaced with the next instantaneous configuration, from which we define new $c_{i}^{\alpha}$s.}
\label{fig:Eckart_solution}
\end{figure}

In the Eckart frame  formalism of
Eqs.~\ref{eqn:Eckart omega} and~\ref{eqn:Eckart inertia}, one needs to
define a reference configuration  which fixes  $c_{i}^{\alpha}$ over  time.
These are  the components of  the monomers positions of  the reference
configuration   in  the  initial   internal   coordinate
system.   We  choose   $c_{i}^{\alpha}$  in   three  different   ways:
\textbf{(i)} From  an equilibrium configuration  of a star  polymer at
temperature  $0$~K. \textbf{(ii)} The reference configuration is obtained by Metropolis  Monte  Carlo (MC)  simulation  at  the  desired temperature $T=4$, which enforces by additional terms in the Hamiltonian that the configuration matches the  average gyration  tensor components  at every shear  rate. \textbf{(iii)} The  $c_{i}^{\alpha}$s are not constant.  Instead, they
are changed  after a certain  number of sampled configurations  in the
trajectory.  An  instantaneous configuration  is taken as  a reference
configuration  for  the  following  $\tau_w$  in  time,  i.e.  this
configuration is  used to  evaluate the  angular velocity  of rotation
(using the Eckart  frame formalism) from all  the following trajectory
snapshots  within  the  time  window $\tau_{w}$. Next,  the  first
configuration  following  in  the  trajectory  is  taken  as  the  new
reference  configuration. This  procedure  is thus  repeated from  the
start until the end of the sampled trajectory. We analyze the rotation
of molecules  for different lengths of  the time window and  thus give
the result for  this third characterization of rotation  by the Eckart
frame    formalism   in    the   form    of   $3$-dimensional    plots
(Fig.~\ref{fig:Eckart_solution}).

In all three described definitions of $c_{i}^{\alpha}$s, the unit base
vectors  of the  internal  coordinate  system $\mathbf{f_{1}}$,
$\mathbf{f_{2}}$, and  $\mathbf{f_{3}}$ and the  origin of  the Eckart  frame, defined  by $\mathbf{r}_{cm}$,  are
different  in  every snapshot  of  the  sampled trajectory.  Only  the
reference components $c_{i}^{\alpha}$s  remain constant throughout the
whole  trajectory   in  \textbf{(i)}   and  \textbf{(ii)},   while  in
\textbf{(iii)}  also $c_{i}^{\alpha}$s  change in  time, as  described
above. Molecules rotate in a  flow-gradient plane. Therefore, the only
component of  the  molecules' angular  velocity with non  zero-average
is in  the
neutral  direction and we denote,  ${\bm \omegal}  = \left(\omegal_1, \omegal_2, \omegal_3
\right)$ and ${\bm \omegae} =\left(\omegae_1, \omegae_2, \omegae_3 \right)$, where  indices $1$,  $2$, and $3$ denote  the flow,  gradient, and neutral  direction, respectively.

To determine the optimal way  to define $c_{i}^{\alpha}$s, we plot, in
Fig.~\ref{fig:Eckart_solution},  angular  velocities obtained  by  the
Eckart  formalism using  the definitions  \textbf{(i)}, \textbf{(ii)},
and \textbf{(iii)} for solution of star polymers with $12$ arms of $6$
monomers  (connected  by Hookean  springs).  Plots  for the  melt  are
qualitatively  similar and  are not  shown here.  We observe  that the
approach \textbf{(iii)}, in which $c_{i}^{\alpha}$s change every $\tau
_{w}$, gives the  angular velocity surface that at  the shortest $\tau
_{w}$   corresponds    to   the   standard   approach    (i.e.   using
Eqs.~\ref{eqn:standard   eqn.  omega}   and~\ref{eqn:standard  inertia
  tensor}).  With increasing  $\tau  _{w}$, it  approaches the  values
obtained  by  the  approaches  \textbf{(i)} and  \textbf{(ii)}.  At  a
certain value of $\tau _{w}$, we  observe a sharp crossover in angular
velocity  of polymers  at  very  high shear  rates,  which results  in
qualitatively          different         dependencies          $\omegae/\dot{\gamma}\left(W_{i}\right)$  emerging   only  due   to  the
different reference  frames. A similar  crossover is also  observed in
melts, but  is more  prominent in  solutions. Furthermore,  we observe
that this crossover occurs at higher $\tau _{w}$ for the star polymers
with longer arms.

Importantly, we find that the definitions \textbf{(i)} and \textbf{(ii)} yield basically the same results, which are also similar to the results obtained by the definition \textbf{(iii)} after the crossover. Therefore, in the manuscript, we present only results obtained by the definition \textbf{(ii)}.

\begin{acknowledgments}
J. S. and M. P. acknowledge financial support through grants P1-0002 and  J1-7435 from the Slovenian Research Agency. J. S. acknowledges financial support from Slovene Human Resources Development and Scholarship Fund (186. JR). R. D.-B. acknowledges support from the Spanish government  under national MINECO project FIS2013-47350-C5-1-R. Partial support from COST Action MP1305 is kindly acknowledged.
\end{acknowledgments}

\bibliography{bibliography}

\providecommand{\noopsort}[1]{}\providecommand{\singleletter}[1]{#1}%
\begin{thebibliography}{63}%
\makeatletter
\providecommand \@ifxundefined [1]{%
 \@ifx{#1\undefined}
}%
\providecommand \@ifnum [1]{%
 \ifnum #1\expandafter \@firstoftwo
 \else \expandafter \@secondoftwo
 \fi
}%
\providecommand \@ifx [1]{%
 \ifx #1\expandafter \@firstoftwo
 \else \expandafter \@secondoftwo
 \fi
}%
\providecommand \natexlab [1]{#1}%
\providecommand \enquote  [1]{``#1''}%
\providecommand \bibnamefont  [1]{#1}%
\providecommand \bibfnamefont [1]{#1}%
\providecommand \citenamefont [1]{#1}%
\providecommand \href@noop [0]{\@secondoftwo}%
\providecommand \href [0]{\begingroup \@sanitize@url \@href}%
\providecommand \@href[1]{\@@startlink{#1}\@@href}%
\providecommand \@@href[1]{\endgroup#1\@@endlink}%
\providecommand \@sanitize@url [0]{\catcode `\\12\catcode `\$12\catcode
  `\&12\catcode `\#12\catcode `\^12\catcode `\_12\catcode `\%12\relax}%
\providecommand \@@startlink[1]{}%
\providecommand \@@endlink[0]{}%
\providecommand \url  [0]{\begingroup\@sanitize@url \@url }%
\providecommand \@url [1]{\endgroup\@href {#1}{\urlprefix }}%
\providecommand \urlprefix  [0]{URL }%
\providecommand \Eprint [0]{\href }%
\providecommand \doibase [0]{http://dx.doi.org/}%
\providecommand \selectlanguage [0]{\@gobble}%
\providecommand \bibinfo  [0]{\@secondoftwo}%
\providecommand \bibfield  [0]{\@secondoftwo}%
\providecommand \translation [1]{[#1]}%
\providecommand \BibitemOpen [0]{}%
\providecommand \bibitemStop [0]{}%
\providecommand \bibitemNoStop [0]{.\EOS\space}%
\providecommand \EOS [0]{\spacefactor3000\relax}%
\providecommand \BibitemShut  [1]{\csname bibitem#1\endcsname}%
\let\auto@bib@innerbib\@empty
\bibitem [{\citenamefont {Smith}, \citenamefont {Babcock},\ and\ \citenamefont
  {Chu}(1999)}]{Smith:1999}%
  \BibitemOpen
  \bibfield  {author} {\bibinfo {author} {\bibfnamefont {D.~E.}\ \bibnamefont
  {Smith}}, \bibinfo {author} {\bibfnamefont {H.~P.}\ \bibnamefont {Babcock}},
  \ and\ \bibinfo {author} {\bibfnamefont {S.}~\bibnamefont {Chu}},\ }\href
  {\doibase 10.1126/science.283.5408.1724} {\bibfield  {journal} {\bibinfo
  {journal} {Science}\ }\textbf {\bibinfo {volume} {283}},\ \bibinfo {pages}
  {1724} (\bibinfo {year} {1999})}\BibitemShut {NoStop}%
\bibitem [{\citenamefont {Teixeira}\ \emph {et~al.}(2005)\citenamefont
  {Teixeira}, \citenamefont {Babcock}, \citenamefont {Shaqfeh},\ and\
  \citenamefont {Chu}}]{Teixeira:2005}%
  \BibitemOpen
  \bibfield  {author} {\bibinfo {author} {\bibfnamefont {R.~E.}\ \bibnamefont
  {Teixeira}}, \bibinfo {author} {\bibfnamefont {H.~P.}\ \bibnamefont
  {Babcock}}, \bibinfo {author} {\bibfnamefont {E.~S.~G.}\ \bibnamefont
  {Shaqfeh}}, \ and\ \bibinfo {author} {\bibfnamefont {S.}~\bibnamefont
  {Chu}},\ }\href {\doibase 10.1021/ma048077l} {\bibfield  {journal} {\bibinfo
  {journal} {Macromolecules}\ }\textbf {\bibinfo {volume} {38}},\ \bibinfo
  {pages} {581} (\bibinfo {year} {2005})}\BibitemShut {NoStop}%
\bibitem [{\citenamefont {Huang}\ \emph {et~al.}(2011)\citenamefont {Huang},
  \citenamefont {Sutmann}, \citenamefont {Gompper},\ and\ \citenamefont
  {Winkler}}]{tumbling}%
  \BibitemOpen
  \bibfield  {author} {\bibinfo {author} {\bibfnamefont {C.-C.}\ \bibnamefont
  {Huang}}, \bibinfo {author} {\bibfnamefont {G.}~\bibnamefont {Sutmann}},
  \bibinfo {author} {\bibfnamefont {G.}~\bibnamefont {Gompper}}, \ and\
  \bibinfo {author} {\bibfnamefont {R.~G.}\ \bibnamefont {Winkler}},\
  }\href@noop {} {\bibfield  {journal} {\bibinfo  {journal} {Europhys. Lett.}\
  }\textbf {\bibinfo {volume} {93}},\ \bibinfo {pages} {54004} (\bibinfo {year}
  {2011})}\BibitemShut {NoStop}%
\bibitem [{\citenamefont {Costanzo}\ \emph {et~al.}(2016)\citenamefont
  {Costanzo}, \citenamefont {Huang}, \citenamefont {Ianniruberto},
  \citenamefont {Marrucci}, \citenamefont {Hassager},\ and\ \citenamefont
  {Vlassopoulos}}]{Costanzo:2016}%
  \BibitemOpen
  \bibfield  {author} {\bibinfo {author} {\bibfnamefont {S.}~\bibnamefont
  {Costanzo}}, \bibinfo {author} {\bibfnamefont {Q.}~\bibnamefont {Huang}},
  \bibinfo {author} {\bibfnamefont {G.}~\bibnamefont {Ianniruberto}}, \bibinfo
  {author} {\bibfnamefont {G.}~\bibnamefont {Marrucci}}, \bibinfo {author}
  {\bibfnamefont {O.}~\bibnamefont {Hassager}}, \ and\ \bibinfo {author}
  {\bibfnamefont {D.}~\bibnamefont {Vlassopoulos}},\ }\href@noop {} {\bibfield
  {journal} {\bibinfo  {journal} {Macromolecules}\ }\textbf {\bibinfo {volume}
  {49}},\ \bibinfo {pages} {3925} (\bibinfo {year} {2016})}\BibitemShut
  {NoStop}%
\bibitem [{\citenamefont {Winkler}(2016)}]{Winkler:2016}%
  \BibitemOpen
  \bibfield  {author} {\bibinfo {author} {\bibfnamefont {R.~G.}\ \bibnamefont
  {Winkler}},\ }\href@noop {} {\bibfield  {journal} {\bibinfo  {journal} {Soft
  Matter}\ }\textbf {\bibinfo {volume} {12}},\ \bibinfo {pages} {3737}
  (\bibinfo {year} {2016})}\BibitemShut {NoStop}%
\bibitem [{\citenamefont {Chen}\ \emph {et~al.}(2017)\citenamefont {Chen},
  \citenamefont {Zhang}, \citenamefont {Liu}, \citenamefont {Chen},
  \citenamefont {Li},\ and\ \citenamefont {An}}]{Chen:2017}%
  \BibitemOpen
  \bibfield  {author} {\bibinfo {author} {\bibfnamefont {W.}~\bibnamefont
  {Chen}}, \bibinfo {author} {\bibfnamefont {K.}~\bibnamefont {Zhang}},
  \bibinfo {author} {\bibfnamefont {L.}~\bibnamefont {Liu}}, \bibinfo {author}
  {\bibfnamefont {J.}~\bibnamefont {Chen}}, \bibinfo {author} {\bibfnamefont
  {Y.}~\bibnamefont {Li}}, \ and\ \bibinfo {author} {\bibfnamefont
  {L.}~\bibnamefont {An}},\ }\href@noop {} {\bibfield  {journal} {\bibinfo
  {journal} {Macromolecules}\ }\textbf {\bibinfo {volume} {50}},\ \bibinfo
  {pages} {1236} (\bibinfo {year} {2017})}\BibitemShut {NoStop}%
\bibitem [{\citenamefont {Abkarian}, \citenamefont {Faivre},\ and\
  \citenamefont {Viallat}(2007)}]{Abkarian:2007}%
  \BibitemOpen
  \bibfield  {author} {\bibinfo {author} {\bibfnamefont {M.}~\bibnamefont
  {Abkarian}}, \bibinfo {author} {\bibfnamefont {M.}~\bibnamefont {Faivre}}, \
  and\ \bibinfo {author} {\bibfnamefont {A.}~\bibnamefont {Viallat}},\ }\href
  {\doibase 10.1103/PhysRevLett.98.188302} {\bibfield  {journal} {\bibinfo
  {journal} {Phys. Rev. Lett.}\ }\textbf {\bibinfo {volume} {98}},\ \bibinfo
  {pages} {188302} (\bibinfo {year} {2007})}\BibitemShut {NoStop}%
\bibitem [{\citenamefont {Yazdani}\ and\ \citenamefont
  {Bagchi}(2011)}]{Alireza:2011}%
  \BibitemOpen
  \bibfield  {author} {\bibinfo {author} {\bibfnamefont {A.~Z.~K.}\
  \bibnamefont {Yazdani}}\ and\ \bibinfo {author} {\bibfnamefont
  {P.}~\bibnamefont {Bagchi}},\ }\href {\doibase 10.1103/PhysRevE.84.026314}
  {\bibfield  {journal} {\bibinfo  {journal} {Phys. Rev. E}\ }\textbf {\bibinfo
  {volume} {84}},\ \bibinfo {pages} {026314} (\bibinfo {year}
  {2011})}\BibitemShut {NoStop}%
\bibitem [{\citenamefont {Dodson}\ and\ \citenamefont
  {Dimitrakopoulos}(2010)}]{Dodson:2010}%
  \BibitemOpen
  \bibfield  {author} {\bibinfo {author} {\bibfnamefont {W.~R.}\ \bibnamefont
  {Dodson}}\ and\ \bibinfo {author} {\bibfnamefont {P.}~\bibnamefont
  {Dimitrakopoulos}},\ }\href@noop {} {\bibfield  {journal} {\bibinfo
  {journal} {Biophys. J.}\ }\textbf {\bibinfo {volume} {99}},\ \bibinfo {pages}
  {2906} (\bibinfo {year} {2010})}\BibitemShut {NoStop}%
\bibitem [{\citenamefont {Aust}, \citenamefont {Hess},\ and\ \citenamefont
  {Kr{\"{o}}ger}(2002)}]{Aust:2002}%
  \BibitemOpen
  \bibfield  {author} {\bibinfo {author} {\bibfnamefont {C.}~\bibnamefont
  {Aust}}, \bibinfo {author} {\bibfnamefont {S.}~\bibnamefont {Hess}}, \ and\
  \bibinfo {author} {\bibfnamefont {M.}~\bibnamefont {Kr{\"{o}}ger}},\ }\href
  {\doibase 10.1021/ma020710k} {\bibfield  {journal} {\bibinfo  {journal}
  {Macromolecules}\ }\textbf {\bibinfo {volume} {35}},\ \bibinfo {pages} {8621}
  (\bibinfo {year} {2002})}\BibitemShut {NoStop}%
\bibitem [{\citenamefont {Ripoll}, \citenamefont {Winkler},\ and\ \citenamefont
  {Gompper}(2006)}]{Ripoll:2006}%
  \BibitemOpen
  \bibfield  {author} {\bibinfo {author} {\bibfnamefont {M.}~\bibnamefont
  {Ripoll}}, \bibinfo {author} {\bibfnamefont {R.~G.}\ \bibnamefont {Winkler}},
  \ and\ \bibinfo {author} {\bibfnamefont {G.}~\bibnamefont {Gompper}},\
  }\href@noop {} {\bibfield  {journal} {\bibinfo  {journal} {Phys. Rev. Lett.}\
  }\textbf {\bibinfo {volume} {96}},\ \bibinfo {pages} {188302} (\bibinfo
  {year} {2006})}\BibitemShut {NoStop}%
\bibitem [{\citenamefont {Ripoll}, \citenamefont {Winkler},\ and\ \citenamefont
  {Gompper}(2007)}]{Ripoll:2007}%
  \BibitemOpen
  \bibfield  {author} {\bibinfo {author} {\bibfnamefont {M.}~\bibnamefont
  {Ripoll}}, \bibinfo {author} {\bibfnamefont {R.}~\bibnamefont {Winkler}}, \
  and\ \bibinfo {author} {\bibfnamefont {G.}~\bibnamefont {Gompper}},\
  }\href@noop {} {\bibfield  {journal} {\bibinfo  {journal} {Eur. Phys. J. E}\
  }\textbf {\bibinfo {volume} {23}},\ \bibinfo {pages} {349} (\bibinfo {year}
  {2007})}\BibitemShut {NoStop}%
\bibitem [{\citenamefont {Yan}\ \emph {et~al.}(2016)\citenamefont {Yan},
  \citenamefont {Costanzo}, \citenamefont {Jeong}, \citenamefont {Chang},\ and\
  \citenamefont {Vlassopoulos}}]{Yan:2016}%
  \BibitemOpen
  \bibfield  {author} {\bibinfo {author} {\bibfnamefont {Z.-C.}\ \bibnamefont
  {Yan}}, \bibinfo {author} {\bibfnamefont {S.}~\bibnamefont {Costanzo}},
  \bibinfo {author} {\bibfnamefont {Y.}~\bibnamefont {Jeong}}, \bibinfo
  {author} {\bibfnamefont {T.}~\bibnamefont {Chang}}, \ and\ \bibinfo {author}
  {\bibfnamefont {D.}~\bibnamefont {Vlassopoulos}},\ }\href@noop {} {\bibfield
  {journal} {\bibinfo  {journal} {Macromolecules}\ }\textbf {\bibinfo {volume}
  {49}},\ \bibinfo {pages} {1444} (\bibinfo {year} {2016})}\BibitemShut
  {NoStop}%
\bibitem [{\citenamefont {Hsiao}, \citenamefont {Schroeder},\ and\
  \citenamefont {Sing}(2016)}]{Hsiao:2016}%
  \BibitemOpen
  \bibfield  {author} {\bibinfo {author} {\bibfnamefont {K.-W.}\ \bibnamefont
  {Hsiao}}, \bibinfo {author} {\bibfnamefont {C.~M.}\ \bibnamefont
  {Schroeder}}, \ and\ \bibinfo {author} {\bibfnamefont {C.~E.}\ \bibnamefont
  {Sing}},\ }\href@noop {} {\bibfield  {journal} {\bibinfo  {journal}
  {Macromolecules}\ }\textbf {\bibinfo {volume} {49}},\ \bibinfo {pages} {1961}
  (\bibinfo {year} {2016})}\BibitemShut {NoStop}%
\bibitem [{\citenamefont {Yoon}, \citenamefont {Kim},\ and\ \citenamefont
  {Baig}(2016)}]{Yoon:2016}%
  \BibitemOpen
  \bibfield  {author} {\bibinfo {author} {\bibfnamefont {J.}~\bibnamefont
  {Yoon}}, \bibinfo {author} {\bibfnamefont {J.}~\bibnamefont {Kim}}, \ and\
  \bibinfo {author} {\bibfnamefont {C.}~\bibnamefont {Baig}},\ }\href@noop {}
  {\bibfield  {journal} {\bibinfo  {journal} {J. Rheol.}\ }\textbf {\bibinfo
  {volume} {60}},\ \bibinfo {pages} {673} (\bibinfo {year} {2016})}\BibitemShut
  {NoStop}%
\bibitem [{\citenamefont {Chen}, \citenamefont {Chen},\ and\ \citenamefont
  {An}(2013)}]{Chen:2013}%
  \BibitemOpen
  \bibfield  {author} {\bibinfo {author} {\bibfnamefont {W.}~\bibnamefont
  {Chen}}, \bibinfo {author} {\bibfnamefont {J.}~\bibnamefont {Chen}}, \ and\
  \bibinfo {author} {\bibfnamefont {L.}~\bibnamefont {An}},\ }\href {\doibase
  10.1039/C3SM50352F} {\bibfield  {journal} {\bibinfo  {journal} {Soft Matter}\
  }\textbf {\bibinfo {volume} {9}},\ \bibinfo {pages} {4312} (\bibinfo {year}
  {2013})}\BibitemShut {NoStop}%
\bibitem [{\citenamefont {Chen}\ \emph
  {et~al.}(2015{\natexlab{a}})\citenamefont {Chen}, \citenamefont {Zhao},
  \citenamefont {Liu}, \citenamefont {Chen}, \citenamefont {Li},\ and\
  \citenamefont {An}}]{Chen:2015:1}%
  \BibitemOpen
  \bibfield  {author} {\bibinfo {author} {\bibfnamefont {W.}~\bibnamefont
  {Chen}}, \bibinfo {author} {\bibfnamefont {H.}~\bibnamefont {Zhao}}, \bibinfo
  {author} {\bibfnamefont {L.}~\bibnamefont {Liu}}, \bibinfo {author}
  {\bibfnamefont {J.}~\bibnamefont {Chen}}, \bibinfo {author} {\bibfnamefont
  {Y.}~\bibnamefont {Li}}, \ and\ \bibinfo {author} {\bibfnamefont
  {L.}~\bibnamefont {An}},\ }\href@noop {} {\bibfield  {journal} {\bibinfo
  {journal} {Soft Matter}\ }\textbf {\bibinfo {volume} {11}},\ \bibinfo {pages}
  {5265} (\bibinfo {year} {2015}{\natexlab{a}})}\BibitemShut {NoStop}%
\bibitem [{\citenamefont {Chen}\ \emph
  {et~al.}(2015{\natexlab{b}})\citenamefont {Chen}, \citenamefont {Li},
  \citenamefont {Zhao}, \citenamefont {Liu}, \citenamefont {Chen},\ and\
  \citenamefont {An}}]{Chen:2015:2}%
  \BibitemOpen
  \bibfield  {author} {\bibinfo {author} {\bibfnamefont {W.}~\bibnamefont
  {Chen}}, \bibinfo {author} {\bibfnamefont {Y.}~\bibnamefont {Li}}, \bibinfo
  {author} {\bibfnamefont {H.}~\bibnamefont {Zhao}}, \bibinfo {author}
  {\bibfnamefont {L.}~\bibnamefont {Liu}}, \bibinfo {author} {\bibfnamefont
  {J.}~\bibnamefont {Chen}}, \ and\ \bibinfo {author} {\bibfnamefont
  {L.}~\bibnamefont {An}},\ }\href@noop {} {\bibfield  {journal} {\bibinfo
  {journal} {Polymer}\ }\textbf {\bibinfo {volume} {64}},\ \bibinfo {pages} {93
  } (\bibinfo {year} {2015}{\natexlab{b}})}\BibitemShut {NoStop}%
\bibitem [{\citenamefont {Sabli\'{c}}, \citenamefont {Praprotnik},\ and\
  \citenamefont {Delgado-Buscalioni}(364A)}]{Sablic:2016:1}%
  \BibitemOpen
  \bibfield  {author} {\bibinfo {author} {\bibfnamefont {J.}~\bibnamefont
  {Sabli\'{c}}}, \bibinfo {author} {\bibfnamefont {M.}~\bibnamefont
  {Praprotnik}}, \ and\ \bibinfo {author} {\bibfnamefont {R.}~\bibnamefont
  {Delgado-Buscalioni}},\ }\href {\doibase 10.1039/C7SM00364A} {\bibfield
  {journal} {\bibinfo  {journal} {Soft Matter}\ } (\bibinfo {year} {2017, DOI:
  10.1039/C7SM00364A}),\ 10.1039/C7SM00364A},\ \bibinfo {note} {dOI:
  10.1039/C7SM00364A}\BibitemShut {NoStop}%
\bibitem [{\citenamefont {Jain}\ \emph {et~al.}(2015)\citenamefont {Jain},
  \citenamefont {Sasmal}, \citenamefont {Hartkamp}, \citenamefont {Todd},\ and\
  \citenamefont {Prakash}}]{Jain:2015}%
  \BibitemOpen
  \bibfield  {author} {\bibinfo {author} {\bibfnamefont {A.}~\bibnamefont
  {Jain}}, \bibinfo {author} {\bibfnamefont {C.}~\bibnamefont {Sasmal}},
  \bibinfo {author} {\bibfnamefont {R.}~\bibnamefont {Hartkamp}}, \bibinfo
  {author} {\bibfnamefont {B.}~\bibnamefont {Todd}}, \ and\ \bibinfo {author}
  {\bibfnamefont {J.~R.}\ \bibnamefont {Prakash}},\ }\href@noop {} {\bibfield
  {journal} {\bibinfo  {journal} {Chem. Eng. Sci.}\ }\textbf {\bibinfo {volume}
  {121}},\ \bibinfo {pages} {245 } (\bibinfo {year} {2015})},\ \bibinfo {note}
  {2013 Danckwerts Special Issue on Molecular Modelling in Chemical
  Engineering}\BibitemShut {NoStop}%
\bibitem [{\citenamefont {Cerf}(1969)}]{Cerf:1969}%
  \BibitemOpen
  \bibfield  {author} {\bibinfo {author} {\bibfnamefont {R.}~\bibnamefont
  {Cerf}},\ }\href@noop {} {\bibfield  {journal} {\bibinfo  {journal} {J. Chim.
  Phys.}\ }\textbf {\bibinfo {volume} {68}},\ \bibinfo {pages} {479} (\bibinfo
  {year} {1969})}\BibitemShut {NoStop}%
\bibitem [{\citenamefont {Singh}\ \emph {et~al.}(2012)\citenamefont {Singh},
  \citenamefont {Fedosov}, \citenamefont {Chatterji}, \citenamefont {Winkler},\
  and\ \citenamefont {Gompper}}]{Singh:2012}%
  \BibitemOpen
  \bibfield  {author} {\bibinfo {author} {\bibfnamefont {S.~P.}\ \bibnamefont
  {Singh}}, \bibinfo {author} {\bibfnamefont {D.~A.}\ \bibnamefont {Fedosov}},
  \bibinfo {author} {\bibfnamefont {A.}~\bibnamefont {Chatterji}}, \bibinfo
  {author} {\bibfnamefont {R.~G.}\ \bibnamefont {Winkler}}, \ and\ \bibinfo
  {author} {\bibfnamefont {G.}~\bibnamefont {Gompper}},\ }\href@noop {}
  {\bibfield  {journal} {\bibinfo  {journal} {J. Phys.-Condens. Mat.}\ }\textbf
  {\bibinfo {volume} {24}},\ \bibinfo {pages} {464103} (\bibinfo {year}
  {2012})}\BibitemShut {NoStop}%
\bibitem [{\citenamefont {Singh}\ \emph {et~al.}(2013)\citenamefont {Singh},
  \citenamefont {Chatterji}, \citenamefont {Gompper},\ and\ \citenamefont
  {Winkler}}]{Singh:2013}%
  \BibitemOpen
  \bibfield  {author} {\bibinfo {author} {\bibfnamefont {S.~P.}\ \bibnamefont
  {Singh}}, \bibinfo {author} {\bibfnamefont {A.}~\bibnamefont {Chatterji}},
  \bibinfo {author} {\bibfnamefont {G.}~\bibnamefont {Gompper}}, \ and\
  \bibinfo {author} {\bibfnamefont {R.~G.}\ \bibnamefont {Winkler}},\ }\href
  {\doibase 10.1021/ma401571k} {\bibfield  {journal} {\bibinfo  {journal}
  {Macromolecules}\ }\textbf {\bibinfo {volume} {46}},\ \bibinfo {pages} {8026}
  (\bibinfo {year} {2013})}\BibitemShut {NoStop}%
\bibitem [{\citenamefont {Yamamoto}\ and\ \citenamefont
  {Masaoka}(2015)}]{Yamamoto:2015}%
  \BibitemOpen
  \bibfield  {author} {\bibinfo {author} {\bibfnamefont {T.}~\bibnamefont
  {Yamamoto}}\ and\ \bibinfo {author} {\bibfnamefont {N.}~\bibnamefont
  {Masaoka}},\ }\href {\doibase 10.1007/s00397-014-0817-8} {\bibfield
  {journal} {\bibinfo  {journal} {Rheologica Acta}\ }\textbf {\bibinfo {volume}
  {54}},\ \bibinfo {pages} {139} (\bibinfo {year} {2015})}\BibitemShut
  {NoStop}%
\bibitem [{\citenamefont {Xu}\ and\ \citenamefont {Chen}(2016)}]{Xu:2016}%
  \BibitemOpen
  \bibfield  {author} {\bibinfo {author} {\bibfnamefont {X.}~\bibnamefont
  {Xu}}\ and\ \bibinfo {author} {\bibfnamefont {J.}~\bibnamefont {Chen}},\
  }\href@noop {} {\bibfield  {journal} {\bibinfo  {journal} {J. Chem. Phys.}\
  }\textbf {\bibinfo {volume} {144}},\ \bibinfo {eid} {244905} (\bibinfo {year}
  {2016})}\BibitemShut {NoStop}%
\bibitem [{\citenamefont {Delgado-Buscalioni}, \citenamefont {Sabli{\' c}},\
  and\ \citenamefont {Praprotnik}(2015)}]{DelgadoBuscalioni:2015}%
  \BibitemOpen
  \bibfield  {author} {\bibinfo {author} {\bibfnamefont {R.}~\bibnamefont
  {Delgado-Buscalioni}}, \bibinfo {author} {\bibfnamefont {J.}~\bibnamefont
  {Sabli{\' c}}}, \ and\ \bibinfo {author} {\bibfnamefont {M.}~\bibnamefont
  {Praprotnik}},\ }\href {\doibase 10.1140/epjst/e2015-02415-x} {\bibfield
  {journal} {\bibinfo  {journal} {Eur. Phys. J. Special Topics}\ }\textbf
  {\bibinfo {volume} {224}},\ \bibinfo {pages} {2331} (\bibinfo {year}
  {2015})}\BibitemShut {NoStop}%
\bibitem [{\citenamefont {Sabli\'{c}}, \citenamefont {Praprotnik},\ and\
  \citenamefont {Delgado-Buscalioni}(2016)}]{Sablic:2016}%
  \BibitemOpen
  \bibfield  {author} {\bibinfo {author} {\bibfnamefont {J.}~\bibnamefont
  {Sabli\'{c}}}, \bibinfo {author} {\bibfnamefont {M.}~\bibnamefont
  {Praprotnik}}, \ and\ \bibinfo {author} {\bibfnamefont {R.}~\bibnamefont
  {Delgado-Buscalioni}},\ }\href {\doibase 10.1039/C5SM02604K} {\bibfield
  {journal} {\bibinfo  {journal} {Soft Matter}\ }\textbf {\bibinfo {volume}
  {12}},\ \bibinfo {pages} {2416} (\bibinfo {year} {2016})}\BibitemShut
  {NoStop}%
\bibitem [{\citenamefont {{De Fabritiis}}, \citenamefont {Delgado-Buscalioni},\
  and\ \citenamefont {Coveney}(2006)}]{hmd_prl06}%
  \BibitemOpen
  \bibfield  {author} {\bibinfo {author} {\bibfnamefont {G.}~\bibnamefont {{De
  Fabritiis}}}, \bibinfo {author} {\bibfnamefont {R.}~\bibnamefont
  {Delgado-Buscalioni}}, \ and\ \bibinfo {author} {\bibfnamefont
  {P.}~\bibnamefont {Coveney}},\ }\href@noop {} {\bibfield  {journal} {\bibinfo
   {journal} {Phys. Rev. Lett}\ }\textbf {\bibinfo {volume} {97}},\ \bibinfo
  {pages} {134501} (\bibinfo {year} {2006})}\BibitemShut {NoStop}%
\bibitem [{\citenamefont {Eckart}(1935)}]{Eckart:1935}%
  \BibitemOpen
  \bibfield  {author} {\bibinfo {author} {\bibfnamefont {C.}~\bibnamefont
  {Eckart}},\ }\href {\doibase 10.1103/PhysRev.47.552} {\bibfield  {journal}
  {\bibinfo  {journal} {Phys. Rev.}\ }\textbf {\bibinfo {volume} {47}},\
  \bibinfo {pages} {552} (\bibinfo {year} {1935})}\BibitemShut {NoStop}%
\bibitem [{\citenamefont {Wilson}, \citenamefont {Decius},\ and\ \citenamefont
  {Cross}(1955)}]{Wilson:1955}%
  \BibitemOpen
  \bibfield  {author} {\bibinfo {author} {\bibfnamefont {E.}~\bibnamefont
  {Wilson}}, \bibinfo {author} {\bibfnamefont {J.}~\bibnamefont {Decius}}, \
  and\ \bibinfo {author} {\bibfnamefont {P.}~\bibnamefont {Cross}},\
  }\href@noop {} {\emph {\bibinfo {title} {Molecular Vibrations: The Theory of
  Infrared and Raman Vibrational Spectra}}},\ Dover Books on Chemistry Series\
  (\bibinfo  {publisher} {Dover Publications},\ \bibinfo {year}
  {1955})\BibitemShut {NoStop}%
\bibitem [{\citenamefont {Louck}\ and\ \citenamefont
  {Galbraith}(1976)}]{Louck:1976}%
  \BibitemOpen
  \bibfield  {author} {\bibinfo {author} {\bibfnamefont {J.~D.}\ \bibnamefont
  {Louck}}\ and\ \bibinfo {author} {\bibfnamefont {H.~W.}\ \bibnamefont
  {Galbraith}},\ }\href {\doibase 10.1103/RevModPhys.48.69} {\bibfield
  {journal} {\bibinfo  {journal} {Rev. Mod. Phys.}\ }\textbf {\bibinfo {volume}
  {48}},\ \bibinfo {pages} {69} (\bibinfo {year} {1976})}\BibitemShut {NoStop}%
\bibitem [{\citenamefont {Rhee}\ and\ \citenamefont {Kim}(1997)}]{Rhee:1997}%
  \BibitemOpen
  \bibfield  {author} {\bibinfo {author} {\bibfnamefont {Y.~M.}\ \bibnamefont
  {Rhee}}\ and\ \bibinfo {author} {\bibfnamefont {M.~S.}\ \bibnamefont {Kim}},\
  }\href@noop {} {\bibfield  {journal} {\bibinfo  {journal} {J. Chem. Phys.}\
  }\textbf {\bibinfo {volume} {107}},\ \bibinfo {pages} {1394} (\bibinfo {year}
  {1997})}\BibitemShut {NoStop}%
\bibitem [{\citenamefont {Yanao}\ and\ \citenamefont
  {Takatsuka}(2004)}]{Yanao:2004}%
  \BibitemOpen
  \bibfield  {author} {\bibinfo {author} {\bibfnamefont {T.}~\bibnamefont
  {Yanao}}\ and\ \bibinfo {author} {\bibfnamefont {K.}~\bibnamefont
  {Takatsuka}},\ }\href@noop {} {\bibfield  {journal} {\bibinfo  {journal} {J.
  Chem. Phys.}\ }\textbf {\bibinfo {volume} {120}},\ \bibinfo {pages} {8924}
  (\bibinfo {year} {2004})}\BibitemShut {NoStop}%
\bibitem [{\citenamefont {Jane{\v{z}}i{\v{c}}}, \citenamefont {Praprotnik},\
  and\ \citenamefont {Merzel}(2005)}]{Praprotnik:2005}%
  \BibitemOpen
  \bibfield  {author} {\bibinfo {author} {\bibfnamefont {D.}~\bibnamefont
  {Jane{\v{z}}i{\v{c}}}}, \bibinfo {author} {\bibfnamefont {M.}~\bibnamefont
  {Praprotnik}}, \ and\ \bibinfo {author} {\bibfnamefont {F.}~\bibnamefont
  {Merzel}},\ }\href@noop {} {\bibfield  {journal} {\bibinfo  {journal} {J.
  Chem. Phys.}\ }\textbf {\bibinfo {volume} {122}},\ \bibinfo {eid} {174101}
  (\bibinfo {year} {2005})}\BibitemShut {NoStop}%
\bibitem [{\citenamefont {Praprotnik}\ and\ \citenamefont
  {Jane{\v{z}}i{\v{c}}}(2005{\natexlab{a}})}]{Praprotnik:2005:1}%
  \BibitemOpen
  \bibfield  {author} {\bibinfo {author} {\bibfnamefont {M.}~\bibnamefont
  {Praprotnik}}\ and\ \bibinfo {author} {\bibfnamefont {D.}~\bibnamefont
  {Jane{\v{z}}i{\v{c}}}},\ }\href {\doibase 10.1021/ci050168+} {\bibfield
  {journal} {\bibinfo  {journal} {J. Chem. Inf. Model.}\ }\textbf {\bibinfo
  {volume} {45}},\ \bibinfo {pages} {1571} (\bibinfo {year}
  {2005}{\natexlab{a}})}\BibitemShut {NoStop}%
\bibitem [{\citenamefont {Praprotnik}\ and\ \citenamefont
  {Jane{\v{z}}i{\v{c}}}(2005{\natexlab{b}})}]{Praprotnik:2005:2}%
  \BibitemOpen
  \bibfield  {author} {\bibinfo {author} {\bibfnamefont {M.}~\bibnamefont
  {Praprotnik}}\ and\ \bibinfo {author} {\bibfnamefont {D.}~\bibnamefont
  {Jane{\v{z}}i{\v{c}}}},\ }\href@noop {} {\bibfield  {journal} {\bibinfo
  {journal} {J. Chem. Phys.}\ }\textbf {\bibinfo {volume} {122}},\ \bibinfo
  {eid} {174102} (\bibinfo {year} {2005}{\natexlab{b}})}\BibitemShut {NoStop}%
\bibitem [{\citenamefont {Praprotnik}\ and\ \citenamefont
  {Jane{\v{z}}i{\v{c}}}(2005{\natexlab{c}})}]{Praprotnik:2005:3}%
  \BibitemOpen
  \bibfield  {author} {\bibinfo {author} {\bibfnamefont {M.}~\bibnamefont
  {Praprotnik}}\ and\ \bibinfo {author} {\bibfnamefont {D.}~\bibnamefont
  {Jane{\v{z}}i{\v{c}}}},\ }\href@noop {} {\bibfield  {journal} {\bibinfo
  {journal} {J. Chem. Phys.}\ }\textbf {\bibinfo {volume} {122}},\ \bibinfo
  {eid} {174103} (\bibinfo {year} {2005}{\natexlab{c}})}\BibitemShut {NoStop}%
\bibitem [{\citenamefont {Likos}\ \emph {et~al.}(1998)\citenamefont {Likos},
  \citenamefont {L\"owen}, \citenamefont {Watzlawek}, \citenamefont {Abbas},
  \citenamefont {Jucknischke}, \citenamefont {Allgaier},\ and\ \citenamefont
  {Richter}}]{Likos:1998}%
  \BibitemOpen
  \bibfield  {author} {\bibinfo {author} {\bibfnamefont {C.~N.}\ \bibnamefont
  {Likos}}, \bibinfo {author} {\bibfnamefont {H.}~\bibnamefont {L\"owen}},
  \bibinfo {author} {\bibfnamefont {M.}~\bibnamefont {Watzlawek}}, \bibinfo
  {author} {\bibfnamefont {B.}~\bibnamefont {Abbas}}, \bibinfo {author}
  {\bibfnamefont {O.}~\bibnamefont {Jucknischke}}, \bibinfo {author}
  {\bibfnamefont {J.}~\bibnamefont {Allgaier}}, \ and\ \bibinfo {author}
  {\bibfnamefont {D.}~\bibnamefont {Richter}},\ }\href {\doibase
  10.1103/PhysRevLett.80.4450} {\bibfield  {journal} {\bibinfo  {journal}
  {Phys. Rev. Lett.}\ }\textbf {\bibinfo {volume} {80}},\ \bibinfo {pages}
  {4450} (\bibinfo {year} {1998})}\BibitemShut {NoStop}%
\bibitem [{\citenamefont {Likos}(2001)}]{Likos:2001}%
  \BibitemOpen
  \bibfield  {author} {\bibinfo {author} {\bibfnamefont {C.~N.}\ \bibnamefont
  {Likos}},\ }\href {\doibase http://dx.doi.org/10.1016/S0370-1573(00)00141-1}
  {\bibfield  {journal} {\bibinfo  {journal} {Phys. Rep.}\ }\textbf {\bibinfo
  {volume} {348}},\ \bibinfo {pages} {267 } (\bibinfo {year}
  {2001})}\BibitemShut {NoStop}%
\bibitem [{\citenamefont {Grest}\ \emph {et~al.}(2007)\citenamefont {Grest},
  \citenamefont {Fetters}, \citenamefont {Huang},\ and\ \citenamefont
  {Richter}}]{Grest:2007}%
  \BibitemOpen
  \bibfield  {author} {\bibinfo {author} {\bibfnamefont {G.~S.}\ \bibnamefont
  {Grest}}, \bibinfo {author} {\bibfnamefont {L.~J.}\ \bibnamefont {Fetters}},
  \bibinfo {author} {\bibfnamefont {J.~S.}\ \bibnamefont {Huang}}, \ and\
  \bibinfo {author} {\bibfnamefont {D.}~\bibnamefont {Richter}},\ }\enquote
  {\bibinfo {title} {Star polymers: Experiment, theory, and simulation},}\ in\
  \href {\doibase 10.1002/9780470141533.ch2} {\emph {\bibinfo {booktitle}
  {Advances in Chemical Physics}}}\ (\bibinfo  {publisher} {John Wiley and
  Sons, Inc.},\ \bibinfo {year} {2007})\ pp.\ \bibinfo {pages}
  {67--163}\BibitemShut {NoStop}%
\bibitem [{\citenamefont {Hijon}\ \emph {et~al.}(2010)\citenamefont {Hijon},
  \citenamefont {Espa\~nol}, \citenamefont {Vanden-Eijnden},\ and\
  \citenamefont {Delgado-Buscalioni}}]{Hijon_2010}%
  \BibitemOpen
  \bibfield  {author} {\bibinfo {author} {\bibfnamefont {C.}~\bibnamefont
  {Hijon}}, \bibinfo {author} {\bibfnamefont {P.}~\bibnamefont {Espa\~nol}},
  \bibinfo {author} {\bibfnamefont {E.}~\bibnamefont {Vanden-Eijnden}}, \ and\
  \bibinfo {author} {\bibfnamefont {R.}~\bibnamefont {Delgado-Buscalioni}},\
  }\href@noop {} {\bibfield  {journal} {\bibinfo  {journal} {Faraday Discuss.}\
  }\textbf {\bibinfo {volume} {144}},\ \bibinfo {pages} {301} (\bibinfo {year}
  {2010})}\BibitemShut {NoStop}%
\bibitem [{\citenamefont {Veldhorst}, \citenamefont {Dyre},\ and\ \citenamefont
  {Schr{\o}der}(2015)}]{Veldhorst:2015}%
  \BibitemOpen
  \bibfield  {author} {\bibinfo {author} {\bibfnamefont {A.~A.}\ \bibnamefont
  {Veldhorst}}, \bibinfo {author} {\bibfnamefont {J.~C.}\ \bibnamefont {Dyre}},
  \ and\ \bibinfo {author} {\bibfnamefont {T.~B.}\ \bibnamefont
  {Schr{\o}der}},\ }\href@noop {} {\bibfield  {journal} {\bibinfo  {journal}
  {J. Chem. Phys.}\ }\textbf {\bibinfo {volume} {143}},\ \bibinfo {eid}
  {194503} (\bibinfo {year} {2015})}\BibitemShut {NoStop}%
\bibitem [{\citenamefont {Espa{\~{n}}ol}\ and\ \citenamefont
  {Warren}(1995)}]{DPD_Espanol}%
  \BibitemOpen
  \bibfield  {author} {\bibinfo {author} {\bibfnamefont {P.}~\bibnamefont
  {Espa{\~{n}}ol}}\ and\ \bibinfo {author} {\bibfnamefont {P.}~\bibnamefont
  {Warren}},\ }\href@noop {} {\bibfield  {journal} {\bibinfo  {journal}
  {Europhys. Lett.}\ }\textbf {\bibinfo {volume} {30}},\ \bibinfo {pages} {191}
  (\bibinfo {year} {1995})}\BibitemShut {NoStop}%
\bibitem [{\citenamefont {Soddemann}, \citenamefont {D\"unweg},\ and\
  \citenamefont {Kremer}(2003)}]{DPD_Soddemann}%
  \BibitemOpen
  \bibfield  {author} {\bibinfo {author} {\bibfnamefont {T.}~\bibnamefont
  {Soddemann}}, \bibinfo {author} {\bibfnamefont {B.}~\bibnamefont {D\"unweg}},
  \ and\ \bibinfo {author} {\bibfnamefont {K.}~\bibnamefont {Kremer}},\
  }\href@noop {} {\bibfield  {journal} {\bibinfo  {journal} {Phys. Rev. E}\
  }\textbf {\bibinfo {volume} {68}},\ \bibinfo {pages} {046702} (\bibinfo
  {year} {2003})}\BibitemShut {NoStop}%
\bibitem [{\citenamefont {Lees}\ and\ \citenamefont
  {Edwards}(1972)}]{LeesEdwards}%
  \BibitemOpen
  \bibfield  {author} {\bibinfo {author} {\bibfnamefont {A.~W.}\ \bibnamefont
  {Lees}}\ and\ \bibinfo {author} {\bibfnamefont {S.~F.}\ \bibnamefont
  {Edwards}},\ }\href@noop {} {\bibfield  {journal} {\bibinfo  {journal} {J.
  Phys. C Solid State}\ }\textbf {\bibinfo {volume} {5}},\ \bibinfo {pages}
  {1921} (\bibinfo {year} {1972})}\BibitemShut {NoStop}%
\bibitem [{\citenamefont {Evans}\ and\ \citenamefont {Morriss}(1984)}]{Sllod1}%
  \BibitemOpen
  \bibfield  {author} {\bibinfo {author} {\bibfnamefont {D.~J.}\ \bibnamefont
  {Evans}}\ and\ \bibinfo {author} {\bibfnamefont {G.~P.}\ \bibnamefont
  {Morriss}},\ }\href@noop {} {\bibfield  {journal} {\bibinfo  {journal} {Phys.
  Rev. A}\ }\textbf {\bibinfo {volume} {30}},\ \bibinfo {pages} {1528}
  (\bibinfo {year} {1984})}\BibitemShut {NoStop}%
\bibitem [{\citenamefont {Ladd}(1984)}]{Sllod2}%
  \BibitemOpen
  \bibfield  {author} {\bibinfo {author} {\bibfnamefont {A.~J.}\ \bibnamefont
  {Ladd}},\ }\href@noop {} {\bibfield  {journal} {\bibinfo  {journal} {Mol.
  Phys.}\ }\textbf {\bibinfo {volume} {53}},\ \bibinfo {pages} {459} (\bibinfo
  {year} {1984})}\BibitemShut {NoStop}%
\bibitem [{\citenamefont {Delgado-Buscalioni}(2011)}]{Tools}%
  \BibitemOpen
  \bibfield  {author} {\bibinfo {author} {\bibfnamefont {R.}~\bibnamefont
  {Delgado-Buscalioni}},\ }in\ \href@noop {} {\emph {\bibinfo {booktitle} {{In
  Numerical Analysis of Multiscale Computations}}}},\ \bibinfo {editor} {edited
  by\ \bibinfo {editor} {\bibfnamefont {Y.-H. R.~T.}\ \bibnamefont
  {Bj{\"o}rn~Engquist}, \bibfnamefont {Olof~Runborg}}}\ (\bibinfo  {publisher}
  {Sringer},\ \bibinfo {year} {2011})\BibitemShut {NoStop}%
\bibitem [{\citenamefont {Flekkoy}, \citenamefont {Delgado-Buscalioni},\ and\
  \citenamefont {Coveney}(2005)}]{Flekbus}%
  \BibitemOpen
  \bibfield  {author} {\bibinfo {author} {\bibfnamefont {E.~G.}\ \bibnamefont
  {Flekkoy}}, \bibinfo {author} {\bibfnamefont {R.}~\bibnamefont
  {Delgado-Buscalioni}}, \ and\ \bibinfo {author} {\bibfnamefont {P.~V.}\
  \bibnamefont {Coveney}},\ }\href@noop {} {\bibfield  {journal} {\bibinfo
  {journal} {Phys. Rev. E}\ }\textbf {\bibinfo {volume} {72}},\ \bibinfo
  {pages} {026703} (\bibinfo {year} {2005})}\BibitemShut {NoStop}%
\bibitem [{\citenamefont {Tuckerman}(2010)}]{Tuckerman:2010}%
  \BibitemOpen
  \bibfield  {author} {\bibinfo {author} {\bibfnamefont {M.}~\bibnamefont
  {Tuckerman}},\ }\href@noop {} {\emph {\bibinfo {title} {Statistical
  Mechanics: Theory and Molecular Simulation}}},\ Oxford Graduate Texts\
  (\bibinfo  {publisher} {OUP Oxford},\ \bibinfo {year} {2010})\BibitemShut
  {NoStop}%
\bibitem [{\citenamefont {Jendrejack}, \citenamefont {Graham},\ and\
  \citenamefont {de~Pablo}(2000)}]{Jendrejack:2000}%
  \BibitemOpen
  \bibfield  {author} {\bibinfo {author} {\bibfnamefont {R.~M.}\ \bibnamefont
  {Jendrejack}}, \bibinfo {author} {\bibfnamefont {M.~D.}\ \bibnamefont
  {Graham}}, \ and\ \bibinfo {author} {\bibfnamefont {J.~J.}\ \bibnamefont
  {de~Pablo}},\ }\href@noop {} {\bibfield  {journal} {\bibinfo  {journal} {J.
  Chem. Phys.}\ }\textbf {\bibinfo {volume} {113}},\ \bibinfo {pages} {2894}
  (\bibinfo {year} {2000})}\BibitemShut {NoStop}%
\bibitem [{\citenamefont {Jendrejack}, \citenamefont {de~Pablo},\ and\
  \citenamefont {Graham}(2002)}]{Jendrejack:2002}%
  \BibitemOpen
  \bibfield  {author} {\bibinfo {author} {\bibfnamefont {R.~M.}\ \bibnamefont
  {Jendrejack}}, \bibinfo {author} {\bibfnamefont {J.~J.}\ \bibnamefont
  {de~Pablo}}, \ and\ \bibinfo {author} {\bibfnamefont {M.~D.}\ \bibnamefont
  {Graham}},\ }\href@noop {} {\bibfield  {journal} {\bibinfo  {journal} {J.
  Chem. Phys.}\ }\textbf {\bibinfo {volume} {116}},\ \bibinfo {pages} {7752}
  (\bibinfo {year} {2002})}\BibitemShut {NoStop}%
\bibitem [{\citenamefont {Usabiaga}\ and\ \citenamefont
  {Delgado-Buscalioni}(2011)}]{Usabiaga:2011}%
  \BibitemOpen
  \bibfield  {author} {\bibinfo {author} {\bibfnamefont {F.~B.}\ \bibnamefont
  {Usabiaga}}\ and\ \bibinfo {author} {\bibfnamefont {R.}~\bibnamefont
  {Delgado-Buscalioni}},\ }\href {\doibase 10.1002/mats.201100020} {\bibfield
  {journal} {\bibinfo  {journal} {Macromol. Theor. Simul.}\ }\textbf {\bibinfo
  {volume} {20}},\ \bibinfo {pages} {466} (\bibinfo {year} {2011})}\BibitemShut
  {NoStop}%
\bibitem [{\citenamefont {Doyle}\ and\ \citenamefont
  {Underhill}(2005)}]{Doyle:2005}%
  \BibitemOpen
  \bibfield  {author} {\bibinfo {author} {\bibfnamefont {P.~S.}\ \bibnamefont
  {Doyle}}\ and\ \bibinfo {author} {\bibfnamefont {P.~T.}\ \bibnamefont
  {Underhill}},\ }\enquote {\bibinfo {title} {Brownian dynamics simulations of
  polymers and soft matter},}\ in\ \href {\doibase
  10.1007/978-1-4020-3286-8_140} {\emph {\bibinfo {booktitle} {Handbook of
  Materials Modeling: Methods}}},\ \bibinfo {editor} {edited by\ \bibinfo
  {editor} {\bibfnamefont {S.}~\bibnamefont {Yip}}}\ (\bibinfo  {publisher}
  {Springer Netherlands},\ \bibinfo {address} {Dordrecht},\ \bibinfo {year}
  {2005})\ pp.\ \bibinfo {pages} {2619--2630}\BibitemShut {NoStop}%
\bibitem [{\citenamefont {Yasuda}(2006)}]{Yasuda:2006}%
  \BibitemOpen
  \bibfield  {author} {\bibinfo {author} {\bibfnamefont {K.}~\bibnamefont
  {Yasuda}},\ }\href {\doibase 10.4188/jte.52.171} {\bibfield  {journal}
  {\bibinfo  {journal} {J. Text. Eng.}\ }\textbf {\bibinfo {volume} {52}},\
  \bibinfo {pages} {171} (\bibinfo {year} {2006})}\BibitemShut {NoStop}%
\bibitem [{\citenamefont {Aho}(2011)}]{Aho:2011}%
  \BibitemOpen
  \bibfield  {author} {\bibinfo {author} {\bibfnamefont {J.}~\bibnamefont
  {Aho}},\ }\href@noop {} {\emph {\bibinfo {title} {Rheological
  Characterization of Polymer Melts in Shear and Extension: Measurement
  Reliability and Data for Practical Processing}}}\ (\bibinfo  {publisher}
  {Tampere University of Technology},\ \bibinfo {year} {2011})\BibitemShut
  {NoStop}%
\bibitem [{\citenamefont {Doi}\ and\ \citenamefont {Edwards}(1994)}]{Doi1994}%
  \BibitemOpen
  \bibfield  {author} {\bibinfo {author} {\bibfnamefont {M.}~\bibnamefont
  {Doi}}\ and\ \bibinfo {author} {\bibfnamefont {S.~F.}\ \bibnamefont
  {Edwards}},\ }\href@noop {} {\emph {\bibinfo {title} {The Theory of Polymer
  Dynamics}}}\ (\bibinfo  {publisher} {Clarendon Press - Oxford},\ \bibinfo
  {year} {1994})\BibitemShut {NoStop}%
\bibitem [{\citenamefont {Chremos}\ and\ \citenamefont
  {Douglas}(2015)}]{Chremos:2015}%
  \BibitemOpen
  \bibfield  {author} {\bibinfo {author} {\bibfnamefont {A.}~\bibnamefont
  {Chremos}}\ and\ \bibinfo {author} {\bibfnamefont {J.~F.}\ \bibnamefont
  {Douglas}},\ }\href {\doibase http://dx.doi.org/10.1063/1.4931483} {\bibfield
   {journal} {\bibinfo  {journal} {J. Chem. Phys.}\ }\textbf {\bibinfo {volume}
  {143}},\ \bibinfo {pages} {111104} (\bibinfo {year} {2015})}\BibitemShut
  {NoStop}%
\bibitem [{\citenamefont {Howard}(1937{\natexlab{a}})}]{Howard:1937:1}%
  \BibitemOpen
  \bibfield  {author} {\bibinfo {author} {\bibfnamefont {J.~B.}\ \bibnamefont
  {Howard}},\ }\href@noop {} {\bibfield  {journal} {\bibinfo  {journal} {J.
  Chem. Phys.}\ }\textbf {\bibinfo {volume} {5}},\ \bibinfo {pages} {442}
  (\bibinfo {year} {1937}{\natexlab{a}})}\BibitemShut {NoStop}%
\bibitem [{\citenamefont {Howard}(1937{\natexlab{b}})}]{Howard:1937:2}%
  \BibitemOpen
  \bibfield  {author} {\bibinfo {author} {\bibfnamefont {J.~B.}\ \bibnamefont
  {Howard}},\ }\href@noop {} {\bibfield  {journal} {\bibinfo  {journal} {J.
  Chem. Phys.}\ }\textbf {\bibinfo {volume} {5}},\ \bibinfo {pages} {451}
  (\bibinfo {year} {1937}{\natexlab{b}})}\BibitemShut {NoStop}%
\bibitem [{\citenamefont {Kirtman}(1962)}]{Kirtman:1962}%
  \BibitemOpen
  \bibfield  {author} {\bibinfo {author} {\bibfnamefont {B.}~\bibnamefont
  {Kirtman}},\ }\href@noop {} {\bibfield  {journal} {\bibinfo  {journal} {J.
  Chem. Phys.}\ }\textbf {\bibinfo {volume} {37}},\ \bibinfo {pages} {2516}
  (\bibinfo {year} {1962})}\BibitemShut {NoStop}%
\bibitem [{\citenamefont {Kirtman}(1964)}]{Kirtman:1964}%
  \BibitemOpen
  \bibfield  {author} {\bibinfo {author} {\bibfnamefont {B.}~\bibnamefont
  {Kirtman}},\ }\href@noop {} {\bibfield  {journal} {\bibinfo  {journal} {J.
  Chem. Phys.}\ }\textbf {\bibinfo {volume} {41}},\ \bibinfo {pages} {775}
  (\bibinfo {year} {1964})}\BibitemShut {NoStop}%
\bibitem [{\citenamefont {Kirtman}(1968)}]{Kirtman:1968}%
  \BibitemOpen
  \bibfield  {author} {\bibinfo {author} {\bibfnamefont {B.}~\bibnamefont
  {Kirtman}},\ }\href@noop {} {\bibfield  {journal} {\bibinfo  {journal} {J.
  Chem. Phys.}\ }\textbf {\bibinfo {volume} {49}},\ \bibinfo {pages} {2257}
  (\bibinfo {year} {1968})}\BibitemShut {NoStop}%
\end{thebibliography}%

\end{document}